%% file: main.tex
 \documentclass[final,3p,times,authoryear]{elsarticle}

\usepackage{amssymb}
\usepackage{rotating}
\usepackage{tabularx}
\usepackage{tabularray}
\usepackage{lscape}
\usepackage{booktabs}
\usepackage[flushleft]{threeparttable}
\usepackage{amsmath}
\usepackage{cleveref}
\usepackage{siunitx}
\usepackage{subcaption}
\usepackage{amsmath}
\usepackage{bm}
\usepackage{import}
\usepackage{pifont}
\usepackage{url}

\begin{document}

\begin{frontmatter}



\title{EISim: A Platform for Simulating Intelligent Edge Orchestration Solutions}


\author[1]{Henna Kokkonen\corref{cor1}}\ead{henna.kokkonen@oulu.fi}
\author[1]{Susanna Pirttikangas}\ead{susanna.pirttikangas@oulu.fi}
\author[1]{Lauri Lov{\'e}n}\ead{lauri.loven@oulu.fi}

\affiliation[1]{organization={Center for Ubiquitous Computing, University of Oulu},
            country={Finland}}
\cortext[cor1]{Corresponding author}

\begin{abstract}

To support the stringent requirements of the future intelligent and interactive applications, intelligence needs to become an essential part of the resource management in the edge environment. Developing intelligent orchestration solutions is a challenging and arduous task, where the evaluation and comparison of the proposed solution is a focal point. Simulation is commonly used to evaluate and compare proposed solutions. However, the currently existing, openly available simulators are lacking in terms of supporting the research on intelligent edge orchestration methods. To address this need, this article presents a simulation platform called Edge Intelligence Simulator (EISim), the purpose of which is to facilitate the research on intelligent edge orchestration solutions. EISim is extended from an existing fog simulator called PureEdgeSim. In its current form, EISim supports simulating deep reinforcement learning based solutions and different orchestration control topologies in scenarios related to task offloading and resource pricing on edge. The platform also includes additional tools for creating simulation environments, running simulations for agent training and evaluation, and plotting results.
\end{abstract}




\end{frontmatter}


\section{Introduction}
\label{sec:intro}
\input{Chapters/1_introduction}

\section{Related Work}
\label{sec:relwork}
\input{Chapters/2_related_work}

\section{EISim Implementation}
\label{sec:implementation}
\input{Chapters/3_implementation}

\section{Evaluation}
\label{sec:evaluation}
\input{Chapters/4_evaluation}

\section{Simulation Results and Discussion}
\label{sec:discussion}
\input{Chapters/5_discussion}

\section{Conclusion and Future Work}
\label{sec:conclusion}
\input{Chapters/6_conclusion}


\section*{Software availability}
The source code of EISim is openly available at \url{https://github.com/hennas/EISim}

\section*{CRediT authorship contribution statement}
\textbf{Henna Kokkonen:} Conceptualization, Methodology, Software, Validation, Formal analysis, Investigation, Writing - Original Draft, Visualization.
\textbf{Susanna Pirttikangas:} Resources, Writing - Review \& Editing, Supervision.
\textbf{Lauri Lov{\'e}n:} Conceptualization, Resources, Writing - Review \& Editing, Supervision.

\section*{Declaration of competing interest}
The authors declare that they have no known competing financial interests or personal relationships that could have appeared to influence the work reported in this paper.

\section*{Acknowledgement}
This research has been supported by the Academy of Finland, 6G Flagship program under Grants 346208 and
318927; the ECSEL JU FRACTAL (grant 877056), receiving support from the EU Horizon 2020 programme and
Spain, Italy, Austria, Germany, France, Finland, Switzerland; and finally by Business Finland through the Neural pub/sub research project (diary number 8754/31/2022).

\bibliographystyle{elsarticle-harv} 
\bibliography{citations}





\end{document}

%% file: Chapters/1_introduction.tex
A wide variety of novel, interactive and intelligent applications emerge in areas such as smart city, healthcare and Industry 4.0 (see, e.g., \cite{gilchrist2016, qadri2020}). These applications have high, ever-growing requirements in terms of security, reliability and performance. Currently, the development of these applications is heavily dependent on cloud, the abundant resources of which are a necessity for the computationally intensive Artificial Intelligence (AI) methods. However, cloud-native processing requires transmitting data between the end users and the cloud, which increases the latency, burdens the core network and raises privacy concerns. Hence, several computing paradigms, such as edge and fog computing, Multi-access Edge Computing (MEC) and cloudlets (\cite{Ren2020}), have emerged to bring the computing and storage resources from the cloud to the edge, closer to the end users. Even though these paradigms have differences in their architectural considerations and driving forces, they all have the same essence: placing and using computational resources between the end user and the distant cloud in order to reduce latency and energy consumption, as well as increase security and privacy by keeping the application data local.

Bringing the intelligent applications onto the edge between the end users and the cloud is not a simple task. Traditional AI is inherently centralized and resource consuming, while the edge is inherently distributed and limited in resources. Further, the edge nodes are highly heterogeneous in terms of their capabilities, while the edge environment as a whole is characterized by intermittent connectivity, distributed and non-IID data, as well as geographically distributed, opportunistic computing resources (\cite{Kokkonen2022}). Research on developing and adapting AI methods to the edge environment has been coined as \textit{AI on Edge} (\cite{Loven2019, Deng2019}), which is an active research area with an ample amount of research (\cite{Deng2019, Xu2020, Park2021}). However, in order to fully realize the envisioned interactive and intelligent applications, the orchestration of the edge resources must be intelligent as well.

The research on edge orchestration has been very dispersed as the studies usually focus on the orchestration of only one aspect of the whole environment, such as networks, containers or services (\cite{Kokkonen2022, Taleb2017, de2019network, Zhong2022}). Further, the orchestration paradigms of these aspects typically follow the essence of traditional cloud orchestration, which uses centralized, best-effort and reactive techniques (\cite{Kokkonen2022}). However, the fulfillment of the stringent requirements of the future applications requires that the orchestration enables adaptive, context-aware and autonomous behavior on edge. This requires novel, proactive orchestration solutions that are built upon distributed intelligence. This side of the edge intelligence research, namely developing and applying distributed AI methods for performing orchestration functions, has been coined as \textit{AI for Edge} (\cite{Loven2019, Deng2019}).

The lack of a holistic view on edge orchestration has been a significant deficit in terms of developing novel, distributed orchestration solutions. Only recently there has been efforts to piece together different aspects on orchestration to create a more holistic view. \cite{Kokkonen2022} present an early vision for the future of edge orchestration. The vision relies on a more encompassing view on resources in the computing continuum that spans from the end devices to the cloud. The architecture of the computing continuum is envisioned as a Multi-Agent System (MAS) consisting of nearly autonomous, intelligent, and self-interested agents. Agents manage resources in the computing continuum and aim to fulfill externally set objectives on cost, quality and resource usage. Each agent has local autonomy when it comes to making decisions related to orchestration functions. These nearly autonomous agents form a hierarchy where higher levels control lower levels by setting their objectives and constraints. Such a hierarchy enforces local decision making and follows the idea of loose coupling (\cite{Mammela2021}) with a minimal amount of centralized control.

The vision states that through \textit{AI for Edge}, that is, the development of intelligent solutions for distributed, multi-domain and multi-tenant edge orchestration, the edge environment will eventually evolve into a seamless, autonomous computing continuum. The continuum will be able to orchestrate its limited resources in a globally optimized manner while being aware of and ready to adapt to the dynamic environment. Naturally, realizing such a vision of orchestration built upon MAS paradigm, distributed AI, local autonomy and loose coupling is an extremely challenging endeavor.

The realization of the aforementioned vision brings forth multiple open research questions. Some of the most important questions concern the optimal level of autonomy in the system and the adaptation and application of AI methods in the challenging computing continuum environment. Any proposed solution requires ways to test and evaluate the method and compare it with other potential solutions. 

Simulation is a commonly adopted way for evaluating proposed methods, as it provides a controllable and cost-effective testing environment. However, there is a lack of openly available simulation platforms that would particularly support research on intelligent edge orchestration methods and the level of autonomy in orchestration solutions.

This article presents a simulation platform called \textit{Edge Intelligence Simulator} (EISim), the purpose of which is to facilitate research on intelligent edge orchestration solutions particularly in the context of the aforementioned vision. EISim is built on top of an existing fog simulator called PureEdgeSim (\cite{pureEdgeSim}) (version 5.1.0), extending it towards supporting the easier testing and evaluation of intelligent orchestration methods.

Intelligent orchestration methods particularly refer to (Deep) Reinforcement Learning ((D)RL) based solutions. This is because reinforcement based learning is seen as one key ingredient for intelligent computing continuum orchestration in the vision (\cite{Kokkonen2022}), mainly due to its ability to learn from experience decision-making policies that can adapt to complex systems and achieve long-term optimization. 

In its current form, EISim supports simulating scenarios related to task offloading and resource pricing. Task offloading is the core functionality of PureEdgeSim. EISim adds support for resource pricing, which has a crucial role in the future of edge and fog computing due to the multi-domain nature of the computing continuum environment. The continuum involves many different stakeholders, such as end device owners, infrastructure owners and application providers. The stakeholders offering their resources for the computational demand of other stakeholders naturally want to cover their deployment and operating costs and generate profits.

The focus of EISim is particularly on evaluating and comparing the performance of orchestration solutions against different orchestration control topologies. This is because the control topology of the orchestration solution closely relates to the level of autonomy in the system. Hence, to facilitate the research on the optimal level of autonomy, EISim offers three default task orchestration algorithm implementations that correspond to the three main control topologies, namely \textit{decentralized}, \textit{hybrid} and \textit{centralized}. It is important to note that in its current form, EISim focuses on edge-based processing, meaning that tasks are either processed locally by the edge devices that generate them, or offloaded to the edge servers.

In the decentralized control topology, every device and server in the system is a self-interested, autonomous agent. Each device agent aims to maximize its own utility, while each edge server agent aims to maximize its profit. This control topology has the most autonomous agents, as every server decides independently about the price for task execution on its resources, and devices decide whether to offload their tasks and to which server. 

The hybrid control topology introduces a level of control where edge servers are clustered into groups with assigned cluster heads. A cluster head agent decides the price for task execution in the cluster, and each device agent decides to which cluster it offloads the tasks. The cluster head agent also decides how the offloaded tasks are allocated on the cluster nodes.

Finally, the centralized control topology has the least amount of autonomy, as now there is one central edge server agent that decides the price for the task execution on the platform, as well as the allocation of all offloaded tasks. Device agents only decide whether they offload or not.

Overall, the main contributions of EISim over PureEdgeSim can be summarized as follows:

\begin{itemize}
    \item  EISim adds into PureEdgeSim the capabilities to simulate intelligent, DRL-based solutions and dynamic resource pricing, as well as pre-implemented algorithms for the three main orchestration control topologies.
    \item For investigating dynamic resource pricing methods, EISim allows researchers to plug in their own pricing algorithms for servers. The implemented default algorithm is Deep Deterministic Policy Gradient (DDPG) (\cite{ddpg}), which is a state-of-the-art DRL algorithm for continuous action spaces. 
    \item Compared to PureEdgeSim, EISim offers a more versatile application model, improves the extensibility, and enables the reproducibility of the simulation results.
    \item EISim offers an exclusive set of additional tools for simulation environment setup, agent training and result plotting, which further facilitates the research with EISim. 
\end{itemize}

EISim is validated and evaluated through a large-scale MEC simulation case study that verifies the end-to-end performance of EISim in 24 simulation scenarios. The study demonstrates the capabilities of EISim particularly with regard to training agents and evaluating orchestration solutions against control topologies.

The rest of the article is organized as follows. \Cref{sec:relwork} presents the state of the art in edge and fog simulation. \Cref{sec:implementation} describes the architecture, default implementations, use, and additional tools of EISim. \Cref{sec:evaluation} introduces the simulation case study used to validate and evaluate EISim. \Cref{sec:discussion} analyzes the results of the simulation study and discusses the significance and limitations of EISim. Finally, \Cref{sec:conclusion} concludes the article, as well as charts out the future work.

%% file: Chapters/2_related_work.tex
Many edge and fog simulators have been developed in the research community (\cite{Gill2021simulation, Aral2020simulation}). There does not exist a general purpose edge or fog simulator that could simulate a wide selection of orchestration functions and many details of the edge and fog environments realistically. Instead, each simulator has a specific focus on supporting some subset of functions, and they rely on abstractions to simplify the environment. To which extent different important aspects of the edge environment, such as mobility, network functions, energy consumption, virtualized resources, or possible failures of network nodes and links, are modelled varies widely between the simulators.

The following introduces a selection of existing simulators, specifying their intended use, how they model the environment and what type of orchestration control they support. The simulators were selected based on three factors: 1) the simulator is an edge or fog simulator (it simulates processing on edge or fog nodes), 2) the software of the simulator is open source, and 3) the simulator is well-known (often referenced) or recently published.

iFogSim is one of the most referenced simulators in the literature (\cite{ifogsim, Gill2021simulation}). It builds upon a popular cloud simulator CloudSim (\cite{cloudsim}), and is designed to simulate application placement and scheduling in a fog environment. One application is modelled as a Directed Acyclic Graph (DAG), where the nodes are application components and the edges correspond to data dependencies. The placement policy determines how application components are placed on fog nodes, whereas the scheduling policy determines how the resources of a fog node are divided among the application components placed on it. During simulation, IoT devices generate tuples that are processed by the fog nodes.

iFogSim has several deficiencies. It can only simulate centralized application placement policies. Further, it does not support simulating runtime load-balancing algorithms, does not have any type of mobility model for the devices, nor simulates the possible failures of links and nodes. It only supports tree topologies, meaning that communication between nodes on the same level is not possible. The network model is also very simplistic, because it assigns a fixed latency on each link and does not consider the effect of network load on transmission delays. 

Extensions to iFogSim that aim to address some of the deficiencies have been developed. MobFogSim (\cite{mobfogsim}) adds in mobility support through migration of Virtual Machines (VMs) and containers between cloudlets. It adds two customizable functions to iFogSim: migration policy and migration strategy. Migration policy determines when a user's VM should be migrated. Migration strategy, in turn, determines where the user's VM is migrated and how the migration is performed. MobFogSim also supports customized user mobility patterns that can be given as input data. Besides adding in mobility and migration support, MobFogSim does not address any other deficiencies in iFogSim.

iFogSim2 (\cite{ifogsim2}) redefines many core components of iFogSim and adds in support for mobility, application migration, dynamic cluster formation, and microservice orchestration. It allows implementing customized migration policies, distributed clustering algorithms, as well as microservice placement and scheduling methods. It supports customized mobility patterns that can be read from input files. Further, it allows implementing decentralized runtime load-balancing algorithms, as the node that generates a service request uses the implemented scheduling policy to decide where the request is routed. For finding the placement of microservices on the nodes, iFogSim2 offers a service discovery functionality. However, iFogSim2 does not provide a failure model nor a more realistic network model. 

YAFS (\cite{yafs}) simulates IoT applications in a fog environment. It adopts the application model from iFogSim, describing the dependencies as messages. The network topology is graph based with support for multiple graph formats. YAFS offers three types of customizable policies: selection, placement and population. Whenever a node generates a message, the selection policy decides the receiver node and the route of the message through the network. The placement policy decides the placement of the application modules on the nodes. The population policy decides the placement, message type, and temporal distribution of workload generators. All these policies are dynamic and application-specific, that is, they can be invoked any time during the simulation and each application has its own set of policies. YAFS supports modelling the failures of links and nodes, even though the possible recovery of a failed node is not considered. However, YAFS does not offer any default implementations for mobility models.

EdgeCloudSim (\cite{edgecloudsim}) is another well-known simulator built on top of CloudSim. It allows implementing a centralized edge orchestrator module that decides about the scheduling of user tasks. The tasks are generated by end devices according to a Poisson process. EdgeCloudSim offers some benefits over iFogSim. It offers a mobility model for the end devices, and its network model is more realistic as it takes into account the effect of network load on transmission delays. However, there are many deficiencies. EdgeCloudSim offers only one simple default mobility model. Further, EdgeCloudSim does not model the energy consumption of the end devices and cloud datacenters, nor offers a failure model. EdgeCloudSim also supports only a three-level edge computing environment, where all edge servers are a single hop away from the end devices, and a global cloud is directly above the edge server tier.

PFogSim (\cite{pfogsim}) has been built on top of EdgeCloudSim, extending its features by adding support for simulating multi-layered fog environments, dynamic, multi-hop networking, and the mobility of fog nodes in addition to the end devices. PFogSim has been designed for evaluating service placement methods, and it offers six pre-implemented placement methods. It assumes that each end device has a corresponding application service instance, and during the simulation initialization, the implemented placement method assigns for each device a fog node that hosts its service instance. During simulation, whenever a device generates a task, it is routed through the network to the assigned fog node. PFogSim does not model the energy consumption of the system entities nor the failure of links and nodes. It offers one simple default mobility model. Further, PFogSim does not support implementing customized methods for runtime load-balancing.

Sphere (\cite{sphere}) simulates cloudlet computing environments where small clusters of computing nodes are placed near data sources. The environment is modelled as a directed weighted graph, where the nodes represent clusters and workload generators. Workload is modelled as independent jobs, and one job consists of multiple independent tasks. Sphere allows simulating scheduling strategies on two levels: edge infrastructure level and cluster level. Edge infrastructure level orchestration is responsible for scheduling jobs between clusters, whereas cluster level orchestration is responsible for scheduling tasks on the machines. Sphere does not have any mobility models, because end devices are abstracted as workload generators that represent geographic areas. Further, the failure of a link or a machine is not modelled, nor is the possible runtime load-balancing by migrating the tasks or jobs between machines or clusters.

FogNetSim++ (\cite{fognetsim}) is a fog simulator built on top of OMNeT++ (\cite{omnet}). OMNeT++ is a network simulator that can simulate low-level networking details. Hence, FogNetSim++ inherits the ability to simulate packet- and protocol-level networking details. FogNetSim++ has been developed for simulating scenarios related to task execution on fog nodes and publish / subscribe communication model. Simulations rely on a fog broker that acts as a central orchestrator. The broker manages publishers, subscribers, resource pricing, and handoffs due to mobility. It also provides a communication link with the cloud and schedules tasks on fog servers. The scheduling policy of the broker can be customized. FogNetSim++ offers multiple default mobility models, but does not model the failure of the links or nodes, nor any runtime load-balancing between the fog servers, such as migrating VMs.

PureEdgeSim (\cite{pureEdgeSim}) has been developed for simulating IoT applications in fog and mist computing environments. In contrary to the aforementioned simulators, PureEdgeSim also considers the mist computing aspect, meaning that it simulates task execution on end devices and allows an end device to offload its task to another end device. Applications are modelled as independent tasks, and a customized offloading policy that decides when, where and how the tasks are offloaded can be implemented. The orchestration policy can be run by the cloud, fog nodes or end devices, allowing varying levels of control for the offloading policy. PureEdgeSim offers a more realistic network model for Metropolitan Area Network (MAN) and Wide Area Network (WAN) transmissions, implements one default mobility model, and models node failures due to running out of battery. PureEdgeSim also offers an extensive energy model, because in addition to the energy consumption of computation, the energy consumption in Local Area Network (LAN), MAN, and WAN is measured.

\Cref{tab:simtools} gives a short comparison of the simulators. It states the orchestration functions supported by the simulator, that is, what management policies are customizable by the user. It also shows the application model of the simulator, which can either be graph based or independent tasks/jobs. Graph based simulators model the applications as DAGs, whereas simulators using independent tasks or jobs do not model any application components or dependencies between them. 

The \textit{Control} column in \Cref{tab:simtools} indicates what type of orchestration control topologies are included in the simulator. In other words, it lists the control topologies the implementation of which is facilitated by the simulator. Centralized means that there is only one central orchestrator (e.g., a fog broker) that runs all or a subset of the management policies. Decentralized indicates that some subset of the management policies can be run in a fully autonomous manner by the system entities without any type of centralized control. Finally, hybrid indicates that some subset of the management policies can be run in a manner that is some type of middle form between the centralized and decentralized extremes. For example, the system entities can be clustered into groups, each of which has a controller that runs the management policy inside the group and is able to do its decisions in a nearly or fully autonomous manner. The \textit{Algorithms for different CTs} column indicates whether the simulator explicitly addresses different orchestration controls by offering default implementations for different control topologies. The \textit{DRL support} column indicates whether the simulator has support for DRL agent related workflows.

\begin{table}[!ht]
\begin{center}
  \caption{Edge and fog simulator comparison}
  \label{tab:simtools}
  \scriptsize
  \begin{tabularx}{\linewidth}{
  >{\raggedright\hsize=.6\hsize\linewidth=\hsize}X
  >{\raggedright\hsize=.4\hsize\linewidth=\hsize}X
  >{\centering\hsize=.25\hsize\linewidth=\hsize}X
  >{\centering\hsize=.25\hsize\linewidth=\hsize}X
  >{\centering\hsize=.25\hsize\linewidth=\hsize}X
  >{\centering\arraybackslash\hsize=.25\hsize\linewidth=\hsize}X
  }
    \toprule
    \textbf{Simulator} & \textbf{Supports} & \textbf{App model\textsuperscript{1}} & \textbf{Control\textsuperscript{2}} & \textbf{Algorithms for different CTs\textsuperscript{3}} & \textbf{DRL support}\\
    \midrule
    iFogSim (\cite{ifogsim}) & Placement & G & C &\ding{55} &\ding{55} \\
    \midrule
    MobFogSim (\cite{mobfogsim}) & Placement, migration & G & C &\ding{55} &\ding{55} \\
    \midrule
    iFogSim2 (\cite{ifogsim2}) & Placement, scheduling, migration, clustering & G & C, D &\ding{55} &\ding{55} \\
    \midrule
    YAFS (\cite{yafs}) & Placement, scheduling & G & C, D &\ding{55} &\ding{55} \\
    \midrule
    EdgeCloudSim (\cite{edgecloudsim}) & Scheduling & I & C &\ding{55} &\ding{55} \\
    \midrule
    PFogSim (\cite{pfogsim}) & Placement & I & C &\ding{55} &\ding{55} \\
    \midrule
    Sphere (\cite{sphere}) & Scheduling & I & C, H &\ding{55} &\ding{55} \\
    \midrule
    FogNetSim++ (\cite{fognetsim}) & Scheduling & I & C &\ding{55} &\ding{55} \\
    \midrule
    PureEdgeSim (\cite{pureEdgeSim}) & Offloading & I & C, H, D &\ding{55} &\ding{55} \\
    \midrule
    \textbf{EISim} & Offloading, pricing & I & C, H, D & \ding{51} & \ding{51} \\
    \bottomrule
  \end{tabularx}
  \vspace{1pt} \newline
  \raggedright \ \space \textsuperscript{1)} G = Graph, I = Independent tasks/jobs \newline
  \raggedright \ \space \textsuperscript{2)} C = Centralized, D = Decentralized, H = Hybrid \newline
  \raggedright \ \space \textsuperscript{3)} CTs = Control Topologies
\end{center}
\end{table} 

Based on the review on edge and fog simulators, PureEdgeSim was chosen as a base for EISim because it has been the most actively developed simulator, it supports extensibility well and facilitates the implementation of different control topologies the most. EISim is, as far as is known, the first openly available simulator that specifically supports simulating and comparing different orchestration control topologies and DRL-based orchestration solutions. DRL support includes facilitating hyperparameter tuning, training and evaluation of agents both in single-agent and multi-agent settings. Further, EISim is the first edge simulator to address the dynamic pricing of resources. Some of the simulators do have cost models, most notably FogNetSim++ offers four pricing models, but these are static pricing models where the resource prices are predetermined before simulation.

It is important to note that there exists a simulator called SimEdgeIntel (\cite{SimEdgeIntel}), which can be used to simulate caching algorithms and handover strategies in a MEC environment. It is associated with the word edge intelligence, because the case study in the article uses the tool to simulate a federated DRL algorithm for caching. However, this case study demonstrates that the simulator can be modified by the user to simulate DRL-based solutions. The codebase for the simulator itself does not offer any facilities for DRL, such as support for hyperparameter tuning, training progress monitoring, or pre-implemented DRL solutions. It is also good to note that SimEdgeIntel is not included in the above simulator comparison, because it does not simulate any processing on edge nodes due to its focus on caching. In other words, there is no models for edge servers, applications, or computational capabilities on the edge devices.

%% file: Chapters/3_implementation.tex
Like PureEdgeSim, EISim allows the user to simulate a wide variety of scenarios and deployments. However, EISim has a specific focus on evaluating and comparing the performance of intelligent orchestration solutions against different orchestration control topologies. For this, EISim extends and modifies the core modules of PureEdgeSim, as well as adds new features and modules to PureEdgeSim. The following sections explain the architecture, default implementations and use of EISim, specify the changes made with regard to PureEdgeSim, and present the additional tools that come with EISim.

\subsection{Architecture}\label{sec:eisim_arch}

\begin{figure}[!ht]
\centering
\begin{subfigure}{0.45\textwidth}
  \centering
  \includegraphics[width=\linewidth]{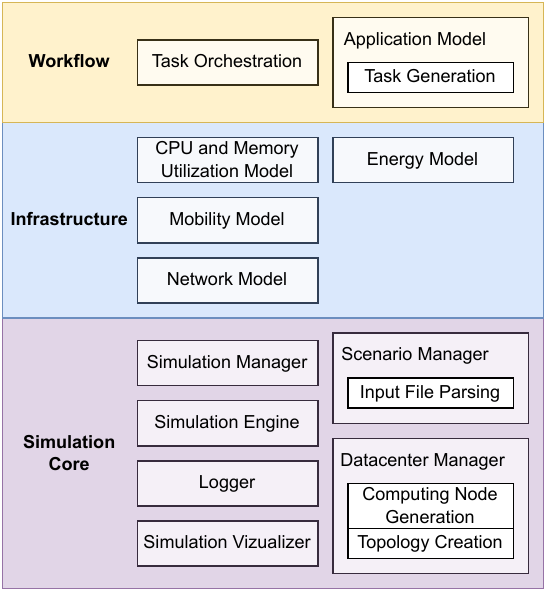}
  \caption{PureEdgeSim}
  \label{fig:pureedge_architecture}
\end{subfigure}
\begin{subfigure}{0.45\textwidth}
  \centering
  \includegraphics[width=\linewidth]{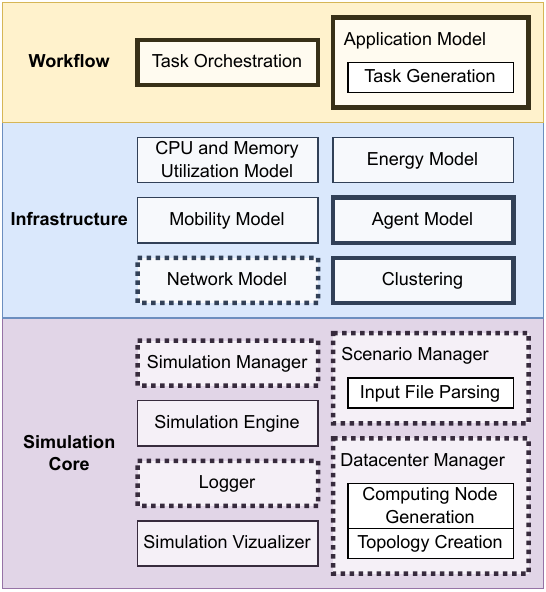}
  \caption{EISim}
  \label{fig:eisim_architecture}
\end{subfigure}
\caption{The architectures of PureEdgeSim and EISim.}
\label{fig:architectures}
\end{figure}

The architecture of PureEdgeSim is shown in \Cref{fig:pureedge_architecture}, and the architecture of EISim is shown in \Cref{fig:eisim_architecture}. PureEdgeSim has a modular, layered architecture, which is retained by EISim. The changes in the EISim architecture compared to PureEdgeSim are shown in \Cref{fig:eisim_architecture}, where thick solid line around a module indicates either that the module is a completely new addition to PureEdgeSim, or that the implementation of the module has been completely changed. Thick dashed line, in turn, indicates that the implementation of the module has been extended for the needs of EISim. 

The modules of PureEdgeSim can be organized into three layers. The lowest layer is the simulation core, which consists of modules that create, manage and monitor the simulation environment. \textit{Simulation Manager} is a central module that initializes the simulation environment, starts and ends the simulation, as well as schedules and handles the main simulation events. It also works as a link between all other modules. \textit{Simulation Engine} module is responsible for running the simulation by managing the event queue. \textit{Logger} records simulation events and calculates the performance metrics at the end of a simulation run. \textit{Simulation Visualizer} module is responsible for creating real-time charts of the simulation map, task success rate, and CPU and network utilization. \textit{Scenario Manager} parses the input files and encapsulates all simulation scenarios. Finally, \textit{Datacenter Manager} module is responsible for creating all the computing nodes (edge devices, edge or fog servers and cloud datacenters) and network links.

The middle layer consists of modules that are responsible for modelling different aspects of the infrastructure. \textit{Network Model} handles all network related events. It manages the data transfers in the network and allocates the bandwidth of each link by taking into account the current network load. \textit{Mobility Model} handles the location and movement of edge devices. \textit{CPU and Memory Utilization Model} handles the resource allocation of a computing node when it receives and executes a task. Finally, \textit{Energy Model} handles the energy consumption of computing nodes and network links.

The highest layer consists of modules that handle the creation and management of workflow resources, specifically tasks. \textit{Application Model} encapsulates the application profiles according to which the tasks are generated. \textit{Task Orchestration} module implements the task orchestration algorithms used by orchestrators to decide on offloading.

The major architectural changes in EISim are the addition of the agent model and clustering modules, and the completely new implementations of the task orchestration and application model modules. To support the new additions and new implementations, other modules had to be extended and modified. The following sections elaborate on the changes.

\subsubsection{Clustering}

The \textit{Clustering} module is responsible for handling edge server clusters. EISim makes it possible to offer cluster information as a part of the edge datacenter specification file. For each edge datacenter, cluster information consists of a non-negative integer that specifies the cluster to which the server belongs, and a boolean value that indicates whether the server is the head of the cluster. During simulation, each edge server is aware of its cluster members, but the cluster members remain static by default.

\subsubsection{Task Orchestration}

EISim implements three default orchestration algorithms that correspond to three main orchestration control topologies. Each of the algorithms makes different assumptions about the edge server clustering. The decentralized orchestration algorithm assumes that each edge server forms a cluster on its own. The centralized orchestration algorithm, in turn, assumes that all edge servers belong to the same cluster with one assigned cluster head. The cluster head functions as a central orchestrator. Finally, the hybrid orchestration algorithm is intended for any type of grouping that resides between the decentralized and centralized extremes.

\begin{figure}[!ht]
\centering
\begin{subfigure}{0.2\textwidth}
  \centering
  \includegraphics[width=\linewidth]{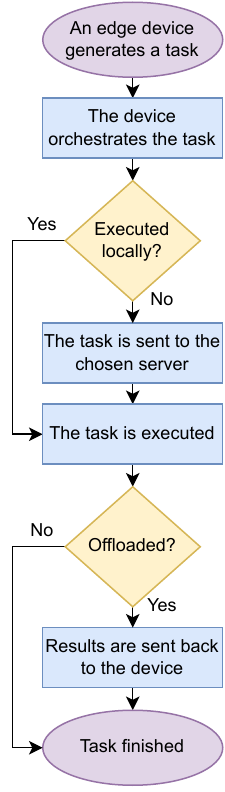}
  \caption{Decentralized}
  \label{fig:D_workflow}
\end{subfigure}
\begin{subfigure}{0.24\textwidth}
  \centering
  \includegraphics[width=\linewidth]{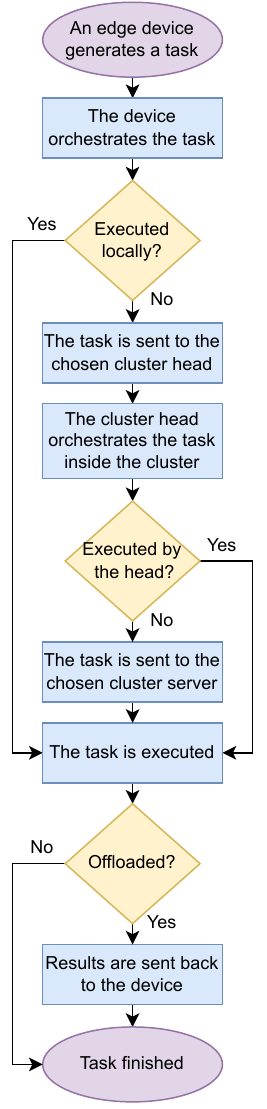}
  \caption{Hybrid}
  \label{fig:H_workflow}
\end{subfigure}
\begin{subfigure}{0.2\textwidth}
  \centering
  \includegraphics[width=\linewidth]{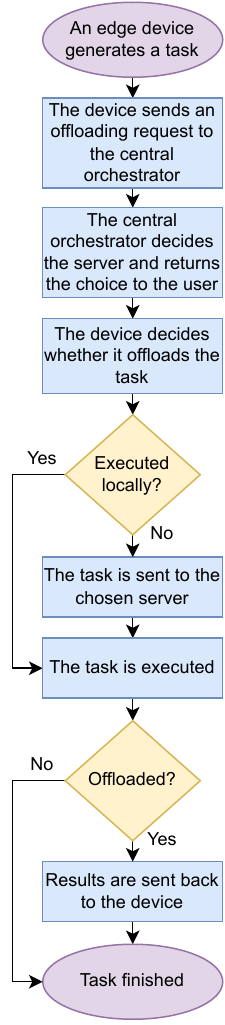}
  \caption{Centralized}
  \label{fig:C_workflow}
\end{subfigure}
\caption{The default task orchestration workflows of each control topology.}
\label{fig:def_workflows}
\end{figure}

Each control topology has its own default orchestration workflow, as shown in \Cref{fig:def_workflows}. \Cref{fig:D_workflow} shows the workflow of the decentralized control topology. Whenever a task is generated, the edge device orchestrates the task by deciding whether to offload and to which server. To make its decision, it uses information from the edge servers, which includes the prices set by the servers. The workflow of the hybrid control topology, as seen in \Cref{fig:H_workflow}, introduces two-phase orchestration, where the edge device first decides whether it offloads and to which cluster. If the device decides to offload, the task is sent to the cluster head, which in turn allocates the task inside the cluster. Finally, in the workflow of the centralized control topology seen in \Cref{fig:C_workflow}, the device first sends an offloading request to the central orchestrator, which chooses the server for executing the task. The choice is returned to the device along with other necessary information. Then, the device makes the final decision whether it offloads or not.

In the centralized control topology, it is important to note that the default implementation does not simulate the sending of the offloading request and its response through the network due to the minuscule sizes of such requests and responses.

More detailed explanations of the default decision making of both sides (edge devices and cluster heads) for each control topology are provided in \Cref{sec:defimpl}.

\subsubsection{Agent Model}

The \textit{Agent Model} module is responsible for handling the training, decision making and monitoring of agents. In the current form of EISim, an agent refers to a pricing agent. In each control topology, cluster heads function as pricing agents that decide a price for task execution on the resources in their cluster. For these agents, the system time is divided into slots and a new price decision is made at the beginning of a slot.

EISim offers a DDPG-based default implementation for a pricing agent, but the users can easily plug in their own implementations. The hyperparameters for pricing agents are given as command-line arguments when starting the simulation. Note that every agent shares the same hyperparameters. More detailed explanations about the possible hyperparameters and the default state space, action space and reward function definitions for each control topology are provided in \Cref{sec:defimpl}.

Each pricing agent in the simulation environment logs its state, price and profit for each slot, as well as calculates the cumulative profit over the simulation run. Further, each pricing agent saves its state at the end of a simulation run, and loads the state at the start of a next one. EISim creates agent- and scenario-specific folders for logging and saving the agent state.

\subsubsection{Application Model}\label{sec:appmodel}

EISim changes the application model of PureEdgeSim into a more versatile one. PureEdgeSim assigns deterministic task generation rates, task lengths and sizes for each application type. This means that every device with the same application type generates identical tasks in terms of input, output and container sizes, and task length in Million Instructions (MIs). To generate more realistic tasks, EISim takes an approach where, for each application type, task generation rates, task lengths and sizes are stochastic with specified expected values.

In EISim, task generation rate refers to the rate of a Poisson process, which is the expected number of task arrivals during a time unit. The task input and container sizes, in turn, are drawn uniformly from a specified range. The output size is determined as a ratio of the task's input size, and this ratio is also drawn uniformly from a specified range. Note that setting the minimum and maximum values of the range to the same value simplifies to the deterministic input, output and container sizes of PureEdgeSim. The task length in MIs is drawn from an exponential distribution with a given expected value. Finally, EISim allows specifying one latency constraint for an application type, which is the same for all the tasks.

In PureEdgeSim, and consequently in EISim, container size is used when registry is enabled for downloading containers. Container size is also used in the default CPU and memory utilization model to specify how much RAM and storage a task uses on a computing node. Input size, in turn, is used when sending a task through the network. In the cases where registry is not used, it may be desirable to set the values of a task's container size and input size equal. This is supported in EISim, as setting the minimum and maximum values of the container size to zero makes EISim use the randomly drawn input size also as a container size.

\subsubsection{Improvements}

EISim modifies the core modules of PureEdgeSim to introduce a set of new improvements. These improvements include the creation of edge nodes that only function as Access Points (APs), the reproducibility of the simulation results, and the better extensibility of the simulator.

APs are specified alongside edge datacenters in an XML file that provides the edge datacenter specifications. Each datacenter element in the file must have a name attribute that contains `dc` if the node is an edge datacenter, and `ap` if the node is an AP. These AP nodes do not have any computational capabilities, and they cannot be a part of the edge server clustering. They only route traffic as a part of the MAN. The MAN links between APs and edge datacenters are defined in the same file.

EISim uses a seed generator that seeds all the random number generators used in the simulation. The user of the simulator can provide a seed for the seed generator through command-line arguments. This allows reproducing the results of a simulation run.

Finally, EISim improves the extensibility of the simulator by allowing users to plug in their own implementation of the computing node generator. This is a new addition to the set of the extensible modules in PureEdgeSim.

\subsection{Default Implementations}\label{sec:defimpl}

For each control topology, EISim implements default decision-making algorithms for price and offloading decisions. These default implementations aim to embrace realism, that is, the algorithms are designed so that they could be potentially deployed under a more practical setting in a large-scale, highly dynamical system. 

The default implementations come with a set of assumptions about the environment. First of all, they assume that each task-generating edge device is its own orchestrator, meaning that it makes the final offloading decision. Second, the only potential task execution locations besides the edge device itself are edge servers. Third, all the edge servers are assumed to belong to the same Edge Service Provider (ESP). Finally, all the edge servers are assumed to be homogeneous in terms of capacity. 

The assumption of homogeneous capacity is justifiable in the simple simulation environment, where edge devices are first located uniformly at random, after which, in case the default mobility model is used, they may move according to randomly drawn mobility and pause durations. Hence, by default, the simulation area does not exhibit distinct areas with different population densities, the existence of which would create a need for placing higher capacity servers to denser areas (\cite{Lahderanta2021}).

\subsubsection{Price Decisions}

In each control topology, the edge platform functions as a time-slotted system, where at the beginning of each slot, the pricing agents decide a price for the task execution. The default slot length is five seconds, and the default pricing scheme is uniform pricing for a task's computational demand. Consequently, a pricing agent sets a price per MI.

The goal of each pricing agent is to maximize the expected long-term profit. Profit is defined as the revenue obtained from the offloading devices minus the processing costs. By default, the processing costs only include the fixed and varying energy costs.

The pricing agents are trained using the DDPG algorithm (\cite{ddpg}). The default structures for the critic and actor networks are as follows. Critic is a feedforward, fully connected network with two hidden layers, each of which has 64 units and ReLU activation. The final activation before output is linear. Actor has the same structure as critic, but the final activation before output is tanh. Both networks are randomly initialized.

As the decision making in the environment is a continuous task without any terminal states, simulation runs can be considered as pseudo-episodes over which the pricing agents can be trained and evaluated. The number of training steps in an episode depends on the price update interval and the total length of the simulation. Agents save their state at the end of a simulation run (episode), and load the state at the beginning of a new one. The agent state in the default DDPG implementation consists of the actor and critic networks, their target counterparts, experience replay, and the state of the exploration noise process.

There are several hyperparameters that are used to control the training process: the size of the experience replay, the size of a mini-batch, discount factor, actor and critic learning rates, parameter for updating the actor and critic target networks, the number of model updates done at the beginning of a new slot, and the parameters for the exploration noise process. Further, EISim allows specifying a number of random decision steps that are done by each pricing agent at the beginning of an episode to improve exploration. A random decision step is a step during which an agent chooses its action uniformly at random. The values for all of these are given as command-line arguments to EISim.

The action space definition is the same for every pricing agent regardless of the control topology. The action space is continuous, consisting of a single real-valued variable between zero and one that corresponds to the price $p_t$ for a slot $t$. The definitions of the state space and reward function depend on the control topology.

In the decentralized control topology, each of the edge servers makes a price decision independently. For one edge server, the state $s_t$ at the beginning of a slot $t$ is defined as $(l_t, \lambda_{t-1})$. Here, $l_t$ is the length of the task queue at the beginning of the slot $t$, and $\lambda_{t-1}$ is the average arrival rate of the tasks in the previous slot $t-1$. The immediate reward for a server is given as

\begin{equation}
     R(s_t, p_t, s_{t+1}) = p_t Q^{MI}_t - \zeta_e \Big ( \tau P_{idle} + (P_{max} - P_{idle}) \frac{Q^{MI}_t}{n^cf} \Big ),
  \label{eq:reward_function}
\end{equation}

where $p_t$ is the price per MI, $Q^{MI}_t$ is the total number of MIs summed over all the tasks that were offloaded to the server in slot $t$, $\zeta_e$ is an energy cost coefficient that defines a cost per joule (J), $\tau$ is the slot length in seconds (s), $P_{idle}$ is the power consumption of the server in Watts (W) when the CPU is idle, $P_{max}$ is the power consumption (W) when the CPU is at 100\%, $n^c$ is the number of cores in the server, and $f$ is the computational capacity of one core in Million Instructions per Second (MIPS). 

In \Cref{eq:reward_function}, $p_t Q^{MI}_t$ is the revenue from the offloading devices, $\zeta_e \tau P_{idle}$ is the fixed energy cost in a slot, and $\zeta_e (P_{max} - P_{idle}) Q^{MI}_t / (n^c f)$ is the varying energy cost that takes into account the excess energy consumption in task processing. The fixed energy cost measures the baseline, load-independent energy cost of the edge server during a slot. The idea of the varying energy cost is to capture the dynamic, load-dependent energy cost. The varying cost calculates how long it takes from the server to process all the arrived tasks with its total computational capacity and multiplies this value with the excess energy consumption.

In the hybrid control topology, each of the cluster heads makes a price decision independently. Now the state for a cluster head is defined as $(l^{avg}_t, \lambda_{t-1})$. Here, $l^{avg}_t$ is the average queue length in the cluster at the beginning of a slot $t$, and $\lambda_{t-1}$ is the average arrival rate of the tasks in the previous slot $t-1$. The reward function for a cluster head is defined as 

\begin{equation}
\begin{split}
    R(s_t, p_t, s_{t+1}) & = p_t Q^{MI}_t - \zeta_e \Big ( |\mathcal{C}| \cdot \tau P_{idle} + |\mathcal{C}| \cdot (P_{max} - P_{idle}) \frac{Q^{MI}_t}{|\mathcal{C}| \cdot n^cf} \Big ) \\
    & = p_t Q^{MI}_t - \zeta_e \Big ( |\mathcal{C}| \cdot \tau P_{idle} + (P_{max} - P_{idle}) \frac{Q^{MI}_t}{n^cf} \Big ),
\end{split}
  \label{eq:reward_functionH}
\end{equation}

where $Q^{MI}_t$ is now the total number of MIs summed over all the tasks that were offloaded to the cluster in slot $t$, and $\mathcal{C}$ is the set of the edge servers in the cluster. Compared to the reward function in the decentralized control topology (\Cref{eq:reward_function}), the fixed energy cost now takes into account the baseline energy consumption of all the edge servers in the cluster. The varying energy cost, in turn, calculates how long it takes to process all the arrived tasks with the cluster's total computational capacity and multiplies this value with the total excess energy consumption of the cluster.

In the centralized control topology, where the central orchestrator is the only pricing agent in the environment, all the servers form one cluster. Hence, the state space and reward function are defined as in the hybrid control topology.

\subsubsection{Task Orchestration Decisions}

Each edge device makes offloading decisions independently based on the information provided by the edge platform. The current default implementation of EISim formulates the offloading problem for each task as a one-shot optimization problem. The decision of an edge device is based on an utility that consists of task execution delay, energy consumption and resource price. 

The default implementation of EISim allows each edge device to be connected to only one AP at a time. After an offloaded task has been executed, the results are sent directly back to the edge device. Note that if the device moves to the coverage area of another AP during offloading, the default implementation of EISim will reroute the task to the new location.

Formally, a task is defined as a tuple $(c, d_{in}, d_{out}, D_{max})$, where $c$ is the computational demand of the task in MIs, $d_{in}$ is the length of the input data in bits, $d_{out}$ is the length of the output data in bits, and $D_{max}$ is the maximum tolerable delay of the task in seconds. EISim sets the task as failed due to delay if the task processing time exceeds $D_{max}$.

The following sections elaborate the edge devices' decision making for each of the control topologies. Further, it is also explained how the cluster heads allocate tasks inside clusters in the hybrid control topology, and how the central orchestrator allocates tasks in the centralized control topology. For the discussions that follow, it is assumed that there are $N$ edge servers in total, and the edge servers are grouped into $K$ clusters in the hybrid control topology. Further, it is assumed that the computing nodes (edge devices and edge servers) use the default CPU utilization model of EISim, which uses First In First Out scheduling and assigns one task to be completely executed by one CPU core.

\subsubsection*{Decentralized}

In the decentralized topology, a device's decision variables form an $N+1$ length vector $\textbf{x}$ indicating the offloading destination (local node + $N$ servers); that is, the decision variable $x_j \in \{0, 1\}$ $\forall j = 0, \dots , N$ and $\sum_j x_j = 1$. The device's utility is defined in terms of minimizing the cost of task execution, which consists of the execution delay, energy consumption and price. Formally, the optimization problem of the edge device can be defined as

\begin{equation}
    \min_{\textbf{x} = [x_0, \dots , x_N]} \sum_{j=0}^{N} x_j (w_d \frac{D_j}{D_{max}} + w_e \frac{E_j}{B_e} + w_p \frac{p_j}{p_{pref}})
\label{eq:user_problem}
\end{equation}
\begin{equation}
    \textrm{s.t.} \;\; x_j \in \{0, 1\} \; \forall j = 0, \dots , N
    \label{eq:user_constraint1}
\end{equation}
\begin{equation}
    \sum_j x_j = 1
    \label{eq:user_constraint2}
\end{equation}
\begin{equation}
    E_j \leq B_e
    \label{eq:user_constraint4}
\end{equation}

Here, $D_j$ is the task execution delay (s), $E_j$ is the energy consumption of the device (J), and $p_j$ is the current price per MI. The values of these depend on the offloading destination, indicated by $x_j$. It is good to note that the price per MI for local node $p_0$ is zero. $w_d$, $w_e$ and $w_p$ are device-specific, normalized weights (i.e., $w_d + w_e + w_p = 1$), which indicate the importance of each factor (delay, energy consumption, price) for the edge device. EISim generates these weights for each device at the start of the simulation. The possible weight values lie on a triangle formed by the points (1,0,0), (0,1,0) and (0,0,1) inside a unit cube, and the weights are generated by sampling a point randomly from the triangle.

$D_{max}$, $B_e$ and $p_{pref}$ are used to normalize the values of $D_j$, $E_j$ and $p_j$, as well as make each quantity of the cost dimensionless. 
$D_{max}$ is the maximum tolerable delay of the task (s), $B_e$ is the battery level of the edge device (J), and $p_{pref}$ quantifies how much the edge device prefers to pay per MI. The default implementation uses $p_{pref} = 0\text{.}01$ for every edge device. 

The constraint in \Cref{eq:user_constraint1} ensures that each decision variable is binary. The second constraint in \Cref{eq:user_constraint2} ensures that only one offloading destination is selected. Finally, the constraint in \Cref{eq:user_constraint4} ensures that the energy consumption does not exceed available energy.

The problem in \Cref{eq:user_problem} is an integer programming problem, which, due to the constraint in \Cref{eq:user_constraint2}, can be solved in linear time ($\mathcal{O}(N)$) by calculating the cost for each $x_j$ and setting $x_j = 1$ for the destination $j$ with the lowest cost.

\textit{Calculating the task execution delay $D_j$.}
When $x_0 = 1$, the offloading destination is the local node itself. The local task execution delay $D_0$ is the sum of the processing delay and the queuing delay at the local node. The processing delay is calculated as $c / f_0$, where $f_0$ is the device's processing capacity per one core in MIPS. The queuing delay is approximated by summing the task lengths of all the tasks currently in the queue and dividing the sum with the total processing capacity of the edge device, which is $n^c_0 f_0$, $n^c_0$ being the number of cores in the device's CPU.

When $x_j = 1$ for some $j \in \{1, \dots , N\}$, the offloading destination is the edge server $j$. The task delay $D_j$ at the edge server $j$ consists of a communication delay and an execution delay. The communication delay is the sum of transmission and propagation delays. The transmission delay from the edge device to an AP is $d_{in} / r^u$, where $r^u$ is the uplink transmission rate. The transmission delay of the task result from an AP to the device is $d_{out} / r^d$, where $r^d$ is the downlink transmission rate. For calculating the propagation delay, the default implementation simply sums the link latencies of the shortest path between the edge device and the server $j$.

The execution delay on a server $j$ is the sum of the processing and queuing delays. The processing delay is calculated as $c / f$. The queuing delay is approximated based on a simple heuristic. Whenever an edge server $j$ handles the price update event, it also calculates an estimate of the queuing time with $Q^{MI}_j / (n^c f)$, where $Q^{MI}_j$ is the total number of MIs summed over all the tasks currently in the queue of the server $j$, and $n^c f$ is the total processing capacity of the server $j$ (same for all $j$). EISim uses this estimate as the queuing delay. The idea here is to mimic the practical situation where such estimate would be announced to the edge devices alongside the price. It gives a crude approximation of the queuing time, but can be considered to be an indication of the queuing delay in the case where the price slot length is short.

\textit{Calculating the energy consumption $E_j$.}
The local energy consumption $E_0$ consists of the energy that the edge device spends on the task execution. This is calculated as $E_0 = P_{max} \: c / (n^c_0 f_0)$. The idea here is to calculate the time it would take from the device to process the task if the whole processing capacity was used, and then multiply the time with the maximum power consumption of the CPU.

The energy consumption $E_j$ when the device offloads the task to an edge server $j$ is the same for all $j$. It consists of the energy the device spends on sending and receiving data. If $P_t$ is the device's transmission power (W) and $P_r$ is the device's receiver power (W), the total energy consumption is calculated as $E_j = P_t \: d_{in} / r^u + P_r \: d_{out} / r^d$.

\subsubsection*{Hybrid}

In the hybrid control topology, a device's decision variables form a $K+1$ length vector $\textbf{x}$ indicating the offloading destination (local node + $K$ clusters). That is, the decision variable $x_k \in \{0, 1\}$ $\forall k = 0, \dots , K$ and $\sum_k x_k = 1$. The device's optimization problem is formulated equivalently to \Cref{eq:user_problem}, but now there are $K+1$ components in $\textbf{x}$ instead of $N+1$. The problem is also solved in the same way by iterating over all $K+1$ options, calculating the cost for each $k$ and setting $x_k = 1$ for the cluster $k$ with the lowest cost.

The energy consumption $E_k \; \forall k = 0, \dots , K$ and the local task execution delay $D_0$ are calculated as explained for the decentralized control topology, because their values depend only on the device's local information. The task execution delay $D_k$ at the cluster $k$ is the sum of the communication and execution delays. The communication delay is calculated as explained for the decentralized topology, but now the calculation of the propagation delay only takes into account the communication distance between the edge device and the head of the cluster $k$. The delays inside the cluster are ignored, because the clusters have been formed based on proximity.

The execution delay inside the cluster is the sum of the processing and queuing delays. The processing delay is calculated as in the decentralized topology. The queuing delay is again approximated with a simple heuristic. Whenever the head of the cluster $k$ handles the price update event, it calculates an estimate of the queuing time with $Q^{MI}_k / (|\mathcal{C}_k| n^c f)$, where $Q^{MI}_k$ is the total number of MIs summed over all the queues and tasks in the cluster $k$, and $\mathcal{C}_k$ is the set of the servers in the cluster $k$. The idea here is to calculate how long it takes from one server inside the cluster to clear its task queue, and then use the average of these times as an estimate of the queuing time. Once again, this is a crude approximation, but it provides an indication of the congestion level inside the cluster, given that the price slot length is short.

When an offloaded task arrives at a cluster head, the default implementation uses a bottom-up strategy (\cite{Loven2022}) to allocate the task. This means that the task is allocated to the server with the lowest workload, which is measured in terms of the task queue length.

\subsubsection*{Centralized}

In the centralized control topology, a device's decision variables form a vector $\textbf{x}$ of length two indicating the offloading destination (local node or the edge platform). As previously, $x_l \in \{0, 1\}$ $\forall l = 0, 1$ and $\sum_l x_l = 1$. The offloading decision reduces to setting $x_1 = 1$ (offload to edge) if $w_d \: D_1 / D_{max} + w_e \: E_1 / B_e + w_p \: p_1 / p_{pref} \leq w_d \: D_0 / D_{max} + w_e \: E_0 / B_e$, otherwise $x_0 = 1$ (process locally). 

The values of $D_0$, $E_0$, and $E_1$ are calculated based on local information. For calculating $D_1$, in practice, the device could send an offloading request to the central orchestrator, informing it about the task characteristics and the device's location. The central orchestrator could calculate the task processing time, estimate propagation delays inside the MAN, and collect queue delay information from the edge servers. Using this information, it could choose the server with the lowest estimated delay. Then it could inform the device about the chosen server, estimated delay, and the price $p_1$, after which the device can decide whether it offloads to the given server or not.
  
As EISim does not simulate the transmission of the offloading requests and responses, the actual default implementation calculates the value of $D_j$ for each edge server $j$ in the same way as explained for the decentralized control topology, and chooses the location that has the lowest estimated cost. The main implementation differences with regard to the decentralized control topology are the use of only one price set by the central orchestrator, and a new event for every edge server that makes them record their queue delay estimate at the beginning of every price slot. The idea here is to mimic the fact that the central orchestrator would collect status information from the edge servers only at the beginning of each price slot to reduce overhead, and then use this information when it allocates tasks during the slot.

\subsection{Use and Extensibility}

EISim is built on Java SE Platform and uses Maven\footnote{\url{https://maven.apache.org/}} as a build automation tool. Running simulations with EISim requires five input files: \textit{cloud.xml}, \textit{edge\_datacenters.xml}, \textit{edge\_devices.xml}, \textit{applications.xml}, and \textit{simulation\_parameters.properties}. The cloud specification file defines cloud datacenters in terms of energy consumption, memory, storage, CPU cores, and processing capacity per core in MIPS. Same information is needed to define edge datacenters in the edge datacenters specification file, but, in addition, the location and cluster must be defined for each edge datacenter, as well as whether a datacenter is peripheral or not. The file for edge datacenters also defines locations for APs. The file must also specify the MAN links between the edge datacenters and the APs.

The edge device specification file must provide specifications for different edge device types. For each type, the percentage of all devices that are of this type must be specified. The other settings provide a way to define a wide variety of different types of edge devices. Edge devices can be mobile or static, and they can be specified to be battery-powered with a given battery capacity and initial battery level. The computing capabilities of edge devices are also specified in terms of memory, storage, CPU cores, and processing capacity per core in MIPS. Further, it must be specified whether an edge device type generates tasks and whether it can act as orchestrator for other edge devices. Finally, the LAN connectivity of an edge device type must be given. Three types are supported, namely cellural, Wi-Fi and ethernet.

The application specification file is used to define different application types according to which tasks are generated. The required parameters for task generation are explained in \Cref{sec:appmodel}. In addition to those, for each application type, the percentage of all task-generating devices that have this type must be specified. Finally, the simulation parameters file must define a wide set of simulation parameters, which can be categorized into general settings (e.g., simulation length), simulation area settings, computing node settings, network settings, and task orchestration settings.

It is important to note that using the default implementation of EISim as is makes certain assumptions about the settings in the simulation parameters file. It is assumed that the value of \textit{enable orchestrators} in the simulation parameters file is set as false, and the only value for \textit{orchestration architectures} is \textit{EDGE ONLY}. This is due to the assumptions made in the default implementation (see \Cref{sec:defimpl}). Further, the value of \textit{orchestration algorithms} can be one of the following: \textit{CENTRALIZED}, \textit{HYBRID}, or \textit{DECENTRALIZED}. It is important to note that only one of these algorithms can be given as input to EISim at a time, because each of them makes different assumptions about the edge server clustering.

In addition to input files, EISim uses command-line arguments to facilitate the setting of the training hyperparameters. Command-line arguments are also used to provide the folder that contains the input files to EISim, as well as to specify an output folder for saving the simulation results, and a model folder for saving the states of the pricing agents. By default, EISim runs in evaluation mode, meaning that the pricing agents in the environment expect to find trained models in the provided model folder. To run the simulation in training mode, it must be turned on with a flag option.

For implementing the deep learning abilities of the pricing agents, EISim uses Deeplearning4j library\footnote{\url{https://deeplearning4j.konduit.ai/}}, which is one of the very few deep learning libraries available for Java. By default, EISim uses the native CPU backend for executing the DNN related computations. This can be easily changed to CUDA GPU backend by changing the value of the \textit{nd4j.backend} property in the Maven project's \textit{pom.xml} file to \textit{nd4j-cuda-X-platform}, where X is the CUDA version. The currently used version of Deeplearning4j (1.0.0-M2.1) supports CUDA versions 11.4 and 11.6.

The \textit{Main} class of EISim is the entry point to the simulator. It parses the command-line arguments and creates the EISim simulation object. The users can modify or replace this class in order to add in their custom implementations. EISim inherits high extensibility from PureEdgeSim, as well as enhances the extensibility by offering a wider set of customizable parts. The simulation object offers eight methods that can be used to set a custom implementation class for mobility model, network model, orchestrator, simulation manager, topology creator, computing node generator, computing node, and task generator. Further, a custom task class can be set through the task generator class, and a custom pricing agent class can be set through the computing node class. EISim contains an abstract class for each customizable part of the simulator, and every provided custom class must inherit from the corresponding abstract class.

\subsection{Additional Tools}

EISim facilitates the research by offering additional tools for environment setup, agent training and result plotting. Even though the simulator itself uses Java programming language, the additional tools for the environment setup and result plotting are made with Python programming language. This is due to Python's simplified syntax and ease of use. Jupyter notebooks\footnote{\url{https://jupyter.org/}} are used as an interactive environment for running the Python codes.

\subsubsection{Environment Setup}

Environment setup consists of three Jupyter notebooks. The first notebook generates a MAN randomly, and this creation process consists of AP placement, topology creation and edge server placement. APs are located on a rectangular simulation area based on covering the area with a hexagonal cell grid. The topology is created with a \textit{Tunable Weight Spanning Tree} (TWST) method proposed by \cite{Soltan2016}. TWST is a low-complexity method for creating a spanning tree between physically placed nodes. It includes a tunable weight, the value of which affects the small-world property of the created spanning tree. Further, EISim also implements a modified version of the link adding procedure presented in the same study (\cite{Soltan2016}), which allows to create more complex and robust network topologies. This procedure includes three tunable weights that affect the small-world and scale-free properties of the resulting network. Finally, the edge servers are co-located with the APs. A given number of edge servers is placed on the created network topology randomly, and the probability of choosing an AP node to host an edge server is proportional to the degree of the AP node.

The second notebook creates the edge server clusters and assigns the cluster heads for the hybrid control topology. An agglomerative hierarchical clustering method is used for grouping the edge servers based on proximity. The default distance measure between a pair of edge servers is the length of the shortest path between the servers in the created topology, where each edge is weighted by the Euclidean distance between the vertices. Finally, inside each cluster, the server with the highest betweenness centrality is chosen as the cluster head.

The final notebook uses the MAN and cluster information saved by the previous two notebooks to automatically create the \textit{edge\_datacenters.xml} setting file. The creation of this file also requires providing the specifications for the edge datacenters. As it is assumed in the current form of EISim that there is only one ESP in the area and that all the edge servers are homogeneous, only one edge server specification needs to be provided. This specification is used for all the edge datacenters in the resulting file.

\subsubsection{Agent Training}

EISim offers a set of bash scripts that can be used as templates for running simulations for hyperparameter tuning, training and evaluation of the pricing agents.

By default, the hyperparameter tuning scripts do a grid search over actor and critic learning rates, testing in total nine combinations. For each combination, the models are trained for 10 rounds with different seeds, after which they are evaluated for five rounds with different seeds. The training scripts, in turn, train the model for 100 rounds with different seeds, plotting the training progress every 20th round. Finally, the evaluation scripts run five evaluation rounds with the trained models, using different seeds.

\subsubsection{Result Plotting}

EISim provides Python codes for plotting the results of each phase of the simulations (hyperparameter tuning, training and evaluation). For hyperparameter tuning and training, the plots focus on the profits of the pricing agents. Explanations and examples of the figures for hyperparameter tuning and training can be seen in \Cref{sec:hypertuning} and \Cref{sec:training}, respectively. For evaluation, the plots focus on comparing the final performance of the system after several evaluation episodes have been run with the final trained models. The main axis of comparison in the plots is the three different control topologies. Averages and confidence intervals of several metrics are plotted. More detailed explanations and examples of the figures for final evaluation can be seen in \Cref{sec:finaleval}.

%% file: Chapters/4_evaluation.tex
To verify the end-to-end performance of EISim and to demonstrate the capabilities of EISim particularly with regard to training agents and evaluating orchestration solutions against control topologies, a simulation case study was conducted. The study focused on a large-scale MEC scenario, where mobile users move in a city area and generate independent tasks. All three control topologies with their associated default pricing and offloading decision-making implementations were simulated on this area. The following sections explain the simulation scenarios, environment and settings used in the simulation study.

\subsection{Simulation Scenarios and Environment}

For each control topology, eight scenarios are simulated, totalling in 24 simulation scenarios. One whole scenario consists of the control topology, edge server count, and user count. For the edge server count, two options are considered. In the first option, the ESP has located 20 high-capacity servers in the city area. In the second option, the ESP has located 100 low-capacity servers in the city area. The interest in choosing these two options is to compare the performance of different control topologies in contrary situations, where there is either only a small number of high-capacity servers or a large number of low-capacity servers. To mimic a large-scale system, the number of mobile users is varied from 1000 to 4000 with a step size of 1000.

The simulation environment is a square area with a side length of 1100 meters. The MAN was created on this area using the environment setup tools of EISim. The AP coverage for placing the APs was set to 45 meters, which resulted in the placement of 247 APs to the area. A tree-topology was created for the APs using the TWST algorithm. For the hybrid control topology, the 20 high-capacity edge servers were clustered into eight groups, and the 100 low-capacity servers were clustered into 19 groups.

\begin{figure}[!ht]
\centering
\begin{subfigure}{0.7\textwidth}
  \centering
  \includegraphics[width=\linewidth]{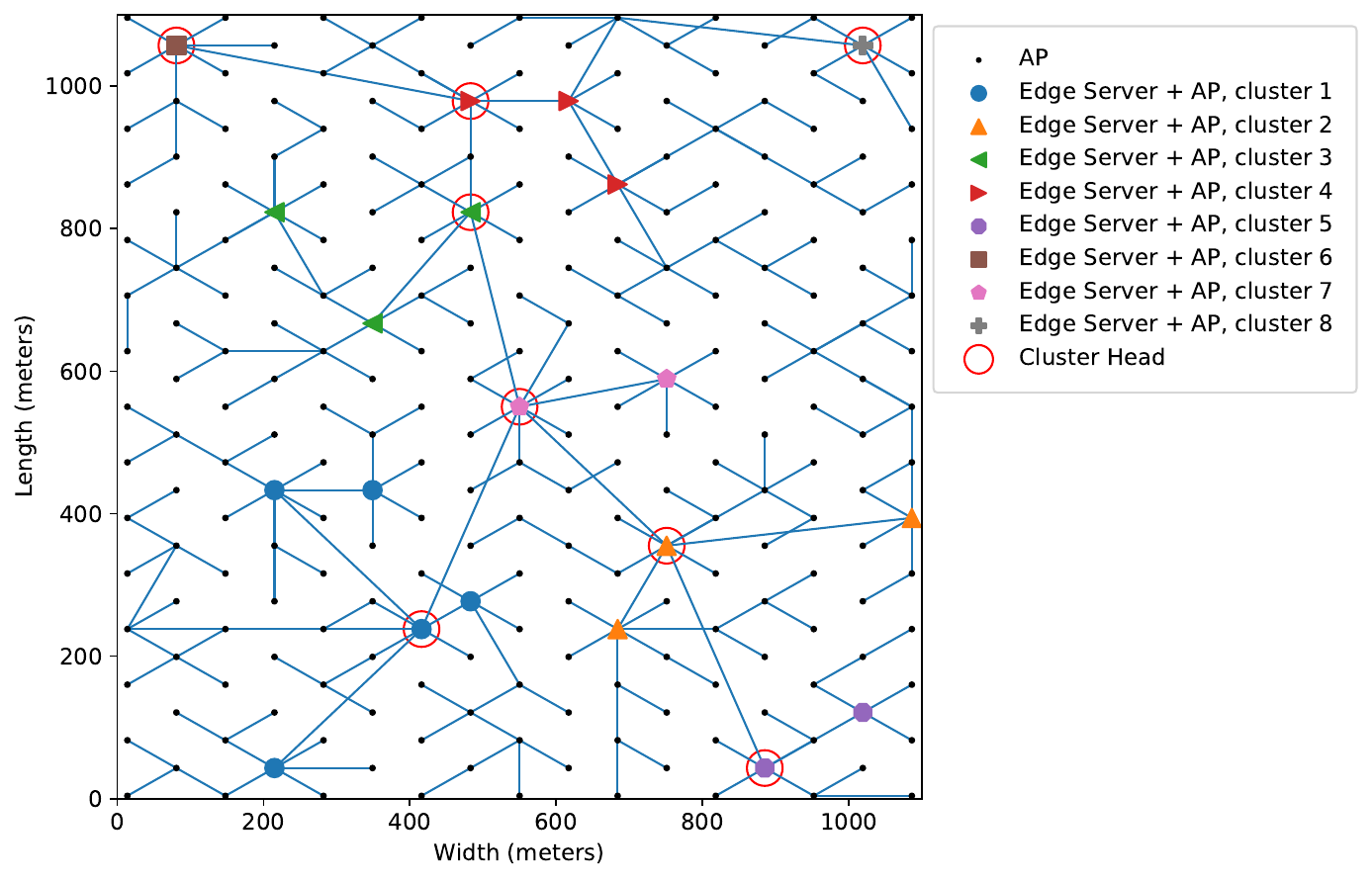}
  \caption{20 edge servers}
  \label{fig:servers_20}
\end{subfigure}
\begin{subfigure}{0.7\textwidth}
  \centering
  \includegraphics[width=\linewidth]{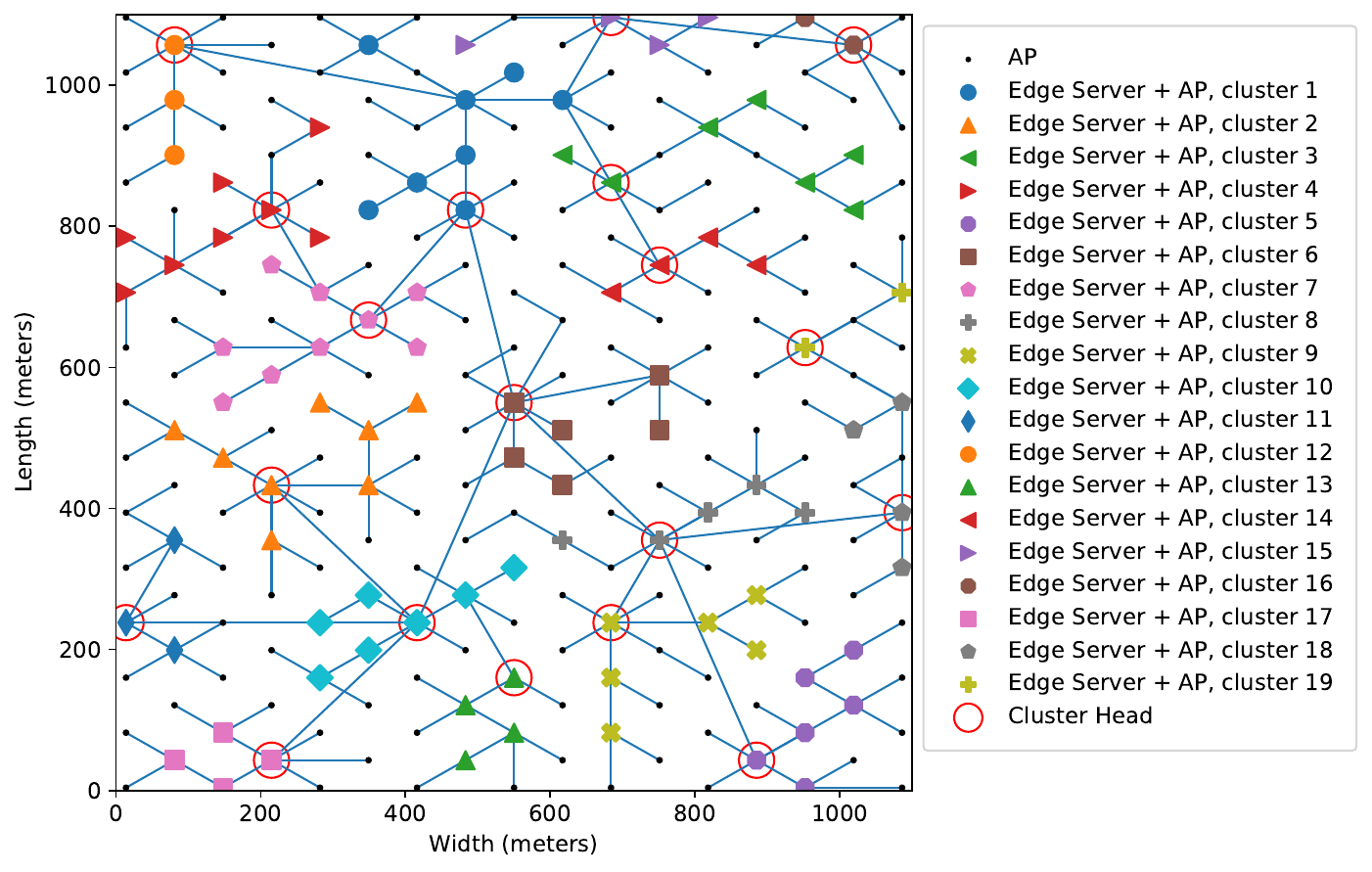}
  \caption{100 edge servers}
  \label{fig:servers_100}
\end{subfigure}
\caption{AP locations, MAN topology and edge server placement in the simulation environment. The edge server clusters and cluster heads for the hybrid control topology are also shown.}
\label{fig:simenvironment}
\end{figure}

The resulting AP placement and MAN topology can be seen in \Cref{fig:simenvironment}. \Cref{fig:servers_20} shows the edge server placement and clustering result for the 20 high-capacity edge servers. \Cref{fig:servers_100}, in turn, shows the edge server placement and clustering result for the 100 low-capacity servers. Even though the created environment is idealized, it can be seen to correspond to a dense deployment of APs in a city centre area.

It is important to note that the central orchestrator for the centralized control topology was chosen in the same way as the cluster heads in the hybrid control topology. In both edge server placement scenarios, the edge server in the middle of the area (see \Cref{fig:simenvironment}) was chosen as the central orchestrator.

\subsection{Specifications and Settings}

\textit{The edge\_datacenters.xml file.} The specifications of the edge servers for the high-capacity and low-capacity server scenarios are shown in \Cref{tab:server_specs}. Note that the specifications are done so that the total MIPS, RAM and storage of all the servers are the same in both scenarios. Here the interest is in examining performance differences when the total capacity in the system is the same, but it is more distributed in the low-capacity server scenario.

\begin{table}[!ht]
\begin{center}
  \caption{Edge server specifications used in EISim evaluation}
  \label{tab:server_specs}
  \scriptsize
  \begin{tabular}{lcc}
    \toprule
    & \textbf{20 high-capacity servers} & \textbf{100 low-capacity servers} \\
    \midrule
    Idle consumption (W) & 105 & 45 \\
    \midrule
    Max consumption (W) & 185 & 95 \\
    \midrule
    Cores & 15 & 6 \\
    \midrule
    MIPS per core & 20,000 & 10,000 \\
    \midrule
    RAM (MB) & 80,000 & 16,000 \\
    \midrule
    Storage (MB) & 1,280,000 & 256,000 \\
    \bottomrule
  \end{tabular}
\end{center}
\end{table}

\textit{The edge\_devices.xml file.} Four edge device types are specified, three of which are mobile. The specifications can be seen in \Cref{tab:device_specs}. Note that the types can be seen to correspond to a higher capacity smartphone, a lower capacity smartphone, a tablet, and a laptop, respectively. All edge device types use Wi-Fi connectivity and are battery-powered.

\begin{table}[!ht]
\begin{center}
  \caption{Mobile device specifications used in EISim evaluation}
  \label{tab:device_specs}
  \scriptsize
  \begin{tabular}{
    >{\raggedright}p{0.2\textwidth}
    >{\centering}p{0.12\textwidth}
    >{\centering}p{0.12\textwidth}
    >{\centering}p{0.12\textwidth}
    >{\centering\arraybackslash}p{0.12\textwidth}
    }
    \toprule
    & \textbf{Device type 1} & \textbf{Device type 2} & \textbf{Device type 3} & \textbf{Device type 4} \\
    \midrule
    Percentage & 30 & 40 & 20 & 10 \\
    \midrule
    Speed (m/s) & 1.1 & 1.1 & 0.6 & 0 \\
    \midrule
    Min pause duration (s) & 60 & 60 & 180 & 0 \\
    \midrule
    Max pause duration (s) & 300 & 300 & 600 & 0 \\
    \midrule
    Min mobility duration (s) & 60 & 60 & 60 & 0 \\
    \midrule
    Max mobility duration (s) & 300 & 300 & 300 & 0 \\
    \midrule
    Battery capacity (Wh) & 19.25 & 15.4 & 25.9 & 56.5 \\
    \midrule
    Idle consumption (W) & 0.9 & 0.6 & 1.1 & 1.7 \\
    \midrule
    Max consumption (W) & 6.2 & 5.5 & 6.5 & 23.6 \\
    \midrule
    Cores & 6 & 4 & 4 & 6 \\
    \midrule
    MIPS per core & 6,000 & 4,000 & 3,000 & 7,000 \\
    \midrule
    RAM (MB) & 6,000 & 4,000 & 2,000 & 8,000 \\
    \midrule
    Storage (MB) & 128,000 & 64,000 & 32,000 & 256,000 \\
    \bottomrule
  \end{tabular}
\end{center}
\end{table}

\textit{The applications.xml file.} All edge devices use the same application type, the specification of which is shown in \Cref{tab:app_types}. Each edge device generates one task per time unit on average. The tasks are computationally demanding on average with a strict latency constraint. The input data size is randomly drawn from a range that varies from small to moderate data size.

\begin{table}[!ht]
\begin{center}
  \caption{Application specification used in EISim evaluation}
  \label{tab:app_types}
  \scriptsize
  \begin{tabular}{ll}
    \toprule
    \textbf{Parameter} & \textbf{Value} \\
    \midrule
    Poisson rate & 1 \\
    \midrule
    Latency (s) & 0.5 \\
    \midrule
    Input size (kB) & U(100, 1000) \\
    \midrule
    Container size (kB) & Equal to input size \\
    \midrule
    Output size (kB) & U(0.2, 0.8) * input size \\
    \midrule
    Task length (MIs) & exp(2000) \\
    \bottomrule
  \end{tabular}
\end{center}
\end{table}

\textit{The simulation\_parameters.properties file.} The values for a selection of important simulation parameters that were the same for all simulation scenarios are shown in \Cref{tab:sim_settings}. The length of one simulation run (episode) is one hour. As the default price slot length is five seconds, there are 720 price updates (training steps) during one episode.

\begin{table}[!ht]
\begin{center}
  \caption{Simulation parameters used in EISim evaluation}
  \label{tab:sim_settings}
  \scriptsize
  \begin{tabular}{
    >{\raggedright}p{0.2\textwidth}
    >{\raggedright}p{0.55\textwidth}
    >{\centering\arraybackslash}p{0.1\textwidth}}
    \toprule
    \textbf{Parameter} & \textbf{Meaning} & \textbf{Value} \\ 
    \midrule
    simulation\_time & Simulation time in minutes & 60 \\
    \midrule
    update\_interval & Mobility and energy consumption update interval for the computing nodes in seconds & 1 \\
    \midrule
    enable\_orchestrators & When enabled, tasks will be sent to a another device/server to make the offloading decisions & false \\
    \midrule
    network\_update\_interval & Transfer update interval for network links in seconds & 1 \\
    \midrule
    man\_bandwidth & MAN link bandwidth in megabits per seconds (Mbps) & 1000 \\
    \midrule
    man\_latency & MAN link latency in seconds & 0.005 \\
    \midrule
    wifi\_bandwidth & Wi-Fi bandwidth in Mbps & 1300 \\
    \midrule
    wifi\_latency & Wi-Fi link latency in seconds & 0.0025 \\
    \midrule
    orchestration\_architectures & Defines which nodes can be considered as offloading destinations in orchestration & EDGE\_ONLY \\
    \bottomrule
  \end{tabular}
\end{center}
\end{table}

%% file: Chapters/5_discussion.tex
\subsection{Hyperparameter Tuning}\label{sec:hypertuning}

\begin{figure}[!ht]
\centering
\includegraphics[width=0.9\linewidth]{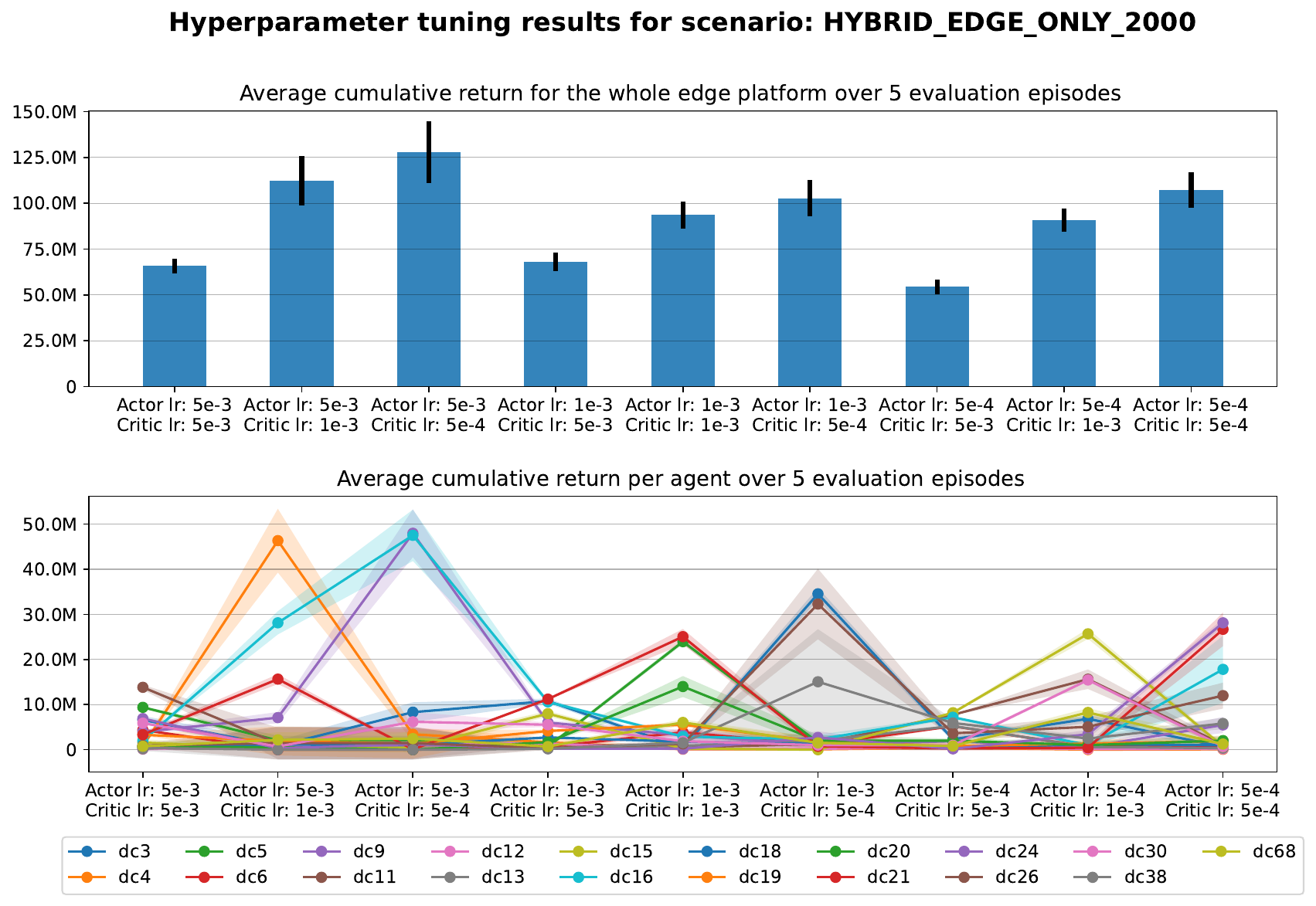}
\caption{An example of a hyperparameter tuning result plot. Results are shown for the hybrid control topology scenario with 100 edge servers and 2000 mobile users.}
\label{fig:htuning_example}
\end{figure}

For each of the 24 simulation scenarios, hyperparameter tuning was done to find the best values for the actor and critic learning rates. To demonstrate how the result plots generated by EISim can be used for selecting the best hyperparameter values, one example of the result plots generated by EISim is shown in \Cref{fig:htuning_example}. It shows the hyperparameter tuning results for the hybrid control topology with 100 servers and 2000 mobile users. The upper plot in \Cref{fig:htuning_example} shows the average cumulative return of the whole edge platform (summed over agents) for each tested hyperparameter combination. The thick black line on top of a bar shows the 95\% confidence interval of the average. The lower plot shows the average cumulative return of each pricing agent for each hyperparameter combination. The shaded area shows the 95\% confidence interval.

In \Cref{fig:htuning_example}, the higher impact of the critic learning rate on the performance is evident. The highest tested learning rate for the critic produces the lowest performance overall and per agent regardless of the value of the actor learning rate. Based on this, the lowest tested learning rate of $\text{5e-4}$ was chosen for the critic. For the actor learning rate, the highest tested value in combination with the critic learning rate of $\text{5e-4}$ produces the highest overall average. However, this result also has the highest degree of uncertainty, as the confidence interval is the widest. Further, when observing the performance of single agents, the high result is based on the good performance of only two agents that are way above the others in the environment. Hence, when taking into account both the overall performance and the performance per agent, the actor learning rate of $\text{5e-4}$ was chosen.

\subsection{Training}\label{sec:training}

\begin{figure}[!ht]
\centering
\includegraphics[width=0.9\linewidth]{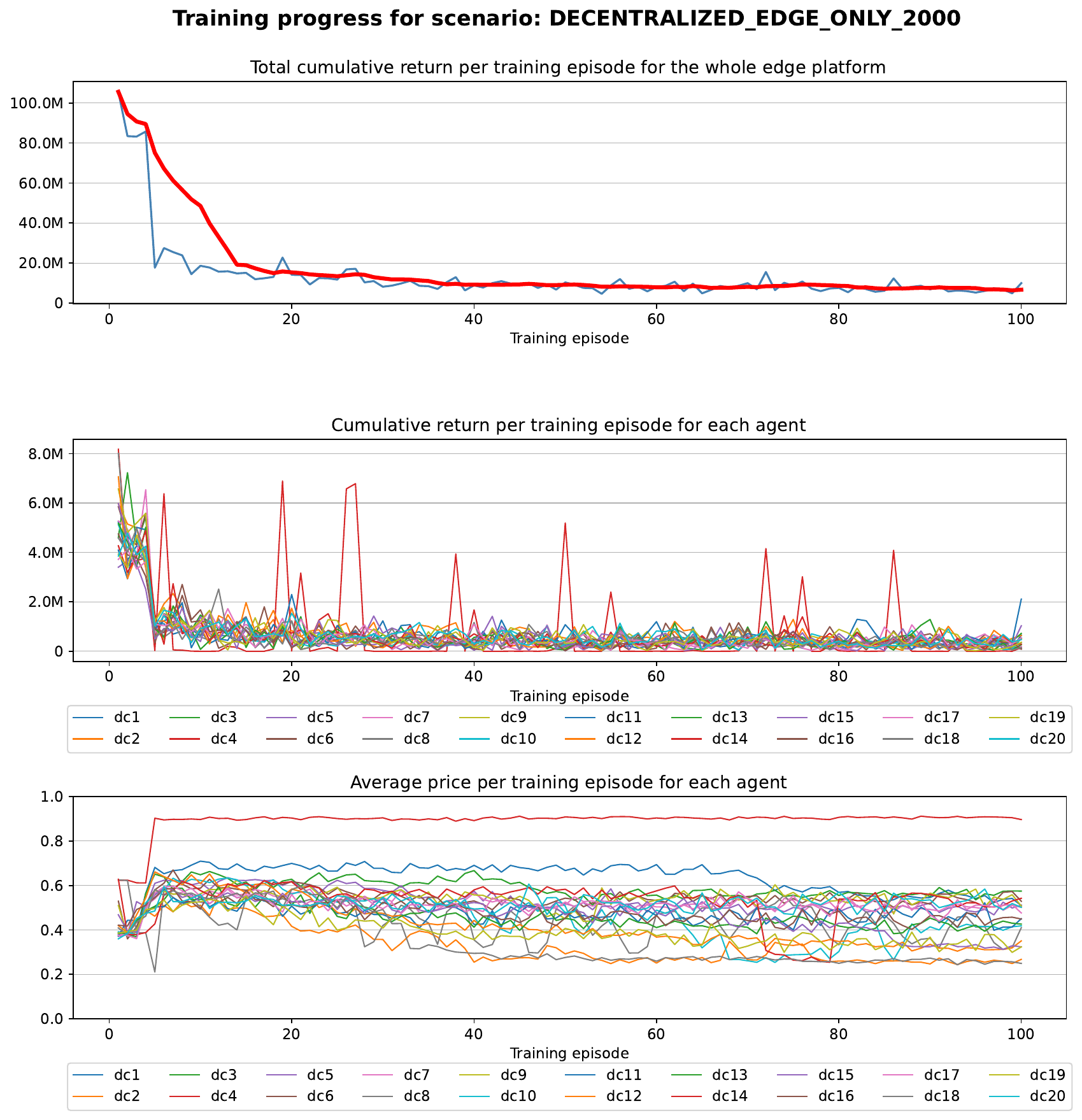}
\caption{An example of a training plot. The training progress is shown for the decentralized control topology scenario with 20 edge servers and 2000 mobile users.}
\label{fig:training_mas_example}
\end{figure}

Every simulation scenario was trained for 100 episodes. To improve exploration, agents chose their action uniformly at random for 500 steps during the first four training episodes, after which only the first price decision was made at random. An example of the training progress plots generated by EISim is shown in \Cref{fig:training_mas_example}. It shows the training progress for the decentralized control topology with 20 servers and 2000 mobile users. The highest plot in \Cref{fig:training_mas_example} shows the total cumulative return per training episode. The thick red line is the simple moving average of the total cumulative return. The middle plot shows the cumulative return of each agent per training episode. The lowest plot shows the average price of each agent per training episode.

It is evident in \Cref{fig:training_mas_example} that using a single-agent algorithm (DDPG) in a multi-agent setting makes it difficult for a single agent to learn an optimal policy due to the inference caused by the other agents. This is a realistic result, as in addition to the reward itself being stochastic, the other agents bring a new source of stochasticity with their evolving policies and action exploration, making it difficult for a single agent to learn the effect of its actions. Further, the agents learn based on the experience saved in the experience replay, but the non-stationarity caused by the other agents means that the dynamics that generated the experience no longer represent the current dynamics for the learners. In other words, the experience can become obsolete very quickly.

\Cref{fig:training_mas_example} shows an interesting strategy for agent `dc4`. It learns very quickly to keep a price level that is way above the average prices of other agents. With this strategy it can occasionally gain a very high return in an episode, as seen in the middle plot. This agent is located at $\text{(215, 823)}$ on the simulation map (see \Cref{fig:servers_20}). It only has one very close competitor in its area, namely `dc14` located at $\text{(349, 667)}$. The agent `dc14` keeps a much lower price level. The examination of its training logs shows that from time to time the exploration noise makes it set the price to zero, which heavily floods the server. 

When taking into account the application profile of the edge devices and the default offloading decision-making logic with the randomly generated importance weights, it is clear that some devices may value low latency much more than low price. These devices may occasionally generate very long tasks, and in case they are in the area of `dc4` while `dc14` is flooded, they may accept its high price in exchange for low-latency execution. This is a perfect example of the destructive effect of the non-stationarity, as the reason why `dc4` thinks that its policy is feasible is a consequence of the exploration noise of `dc14`. The examination of the evaluation logs shows that the agent `dc4` indeed keeps a very high price level constantly, while the agent `dc14` using its learned strategy without any exploration noise keeps a significantly lower price level. Consequently, no edge device offloads to `dc4` during evaluation, as the potential offloaders nearby prefer the low price of `dc14`.

The example above shows that the plots generated by EISim provide good insight into the training progress of the agents. Researchers can use the feedback provided by the plots to develop training methods for the agents. For example, based on the analysis of \Cref{fig:training_mas_example}, it is clear that the training of the agents in the multi-agent setting requires testing and developing interference avoidance techniques. 

\subsection{Performance Evaluation}\label{sec:finaleval}

EISim readily plots multiple metrics that can be used to compare the performance across different control topologies after the agents in the environment have been trained. These metrics are related to the task processing, CPU utilization, energy consumption, network use, and profit. Further, the raw logs of EISim include many more metrics, and additional metrics can be added to the plotted metrics by the user.

EISim generates grouped bar plots for each main scenario and each metric. The main scenario in the conducted simulation study refers to the two server options (20 high-capacity servers or 100 low-capacity servers). In one plot, each control topology has its own element, namely a bar group, and the x-axis values correspond to different edge device counts. The y-axis value for a control topology and edge device count combination is the average of the evaluation metric over the evaluation episodes. The confidence interval of this average is also plotted as a thick line on top of the bar. For the plots shown in this article, the confidence level is 95\%.

Next, plots for some of the evaluation metrics are shown and analyzed. It is exemplified how the plots generated by EISim can be used to compare different control topologies across scenarios. Further, it is shown that EISim is able to output sensible and consistent results.

\begin{figure}[!ht]
\centering
\begin{subfigure}{0.49\textwidth}
  \centering
  \includegraphics[width=\linewidth]{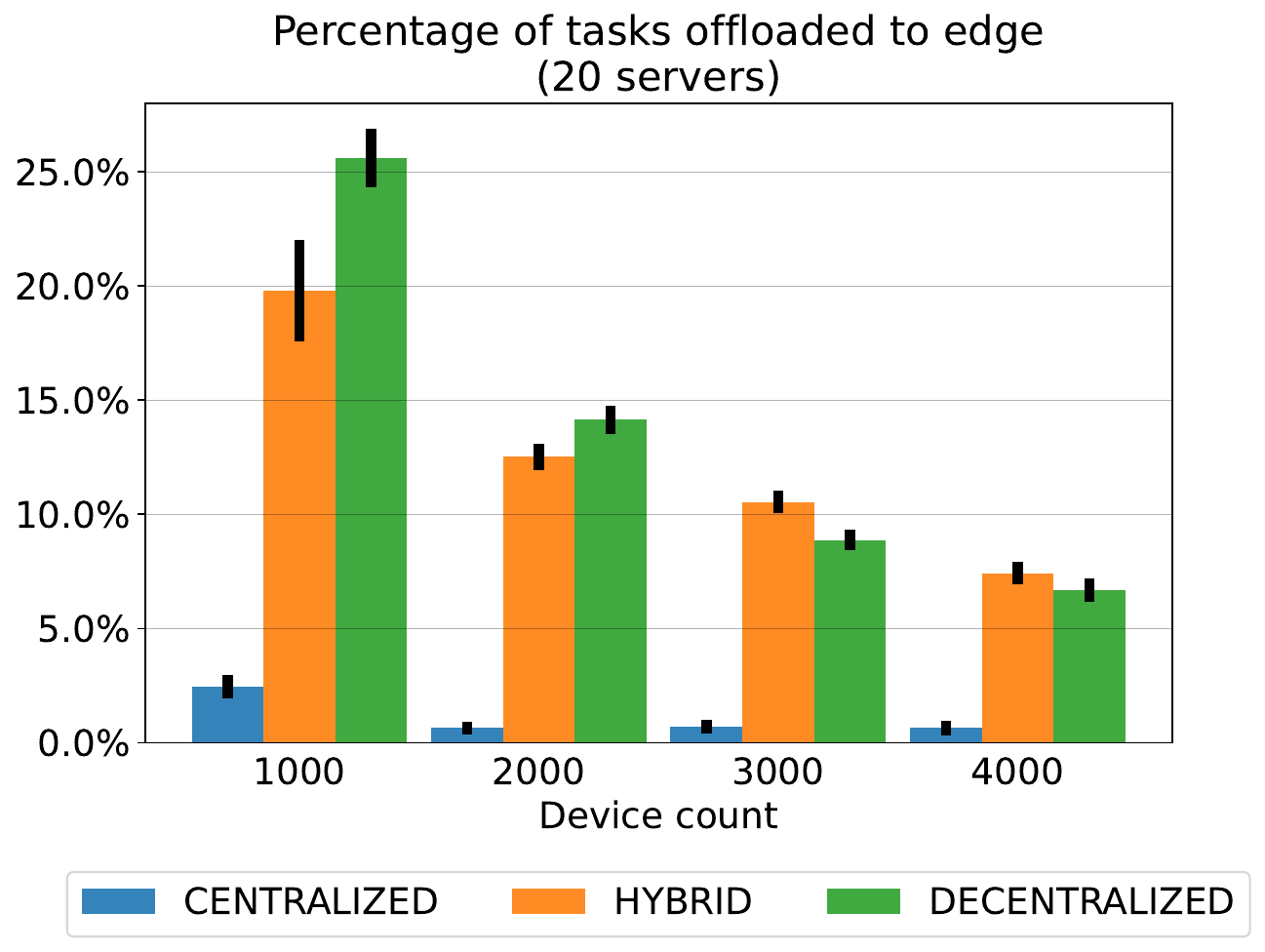}
  \caption{20 servers}
  \label{fig:offloaded20}
\end{subfigure}
\begin{subfigure}{0.49\textwidth}
  \centering
  \includegraphics[width=\linewidth]{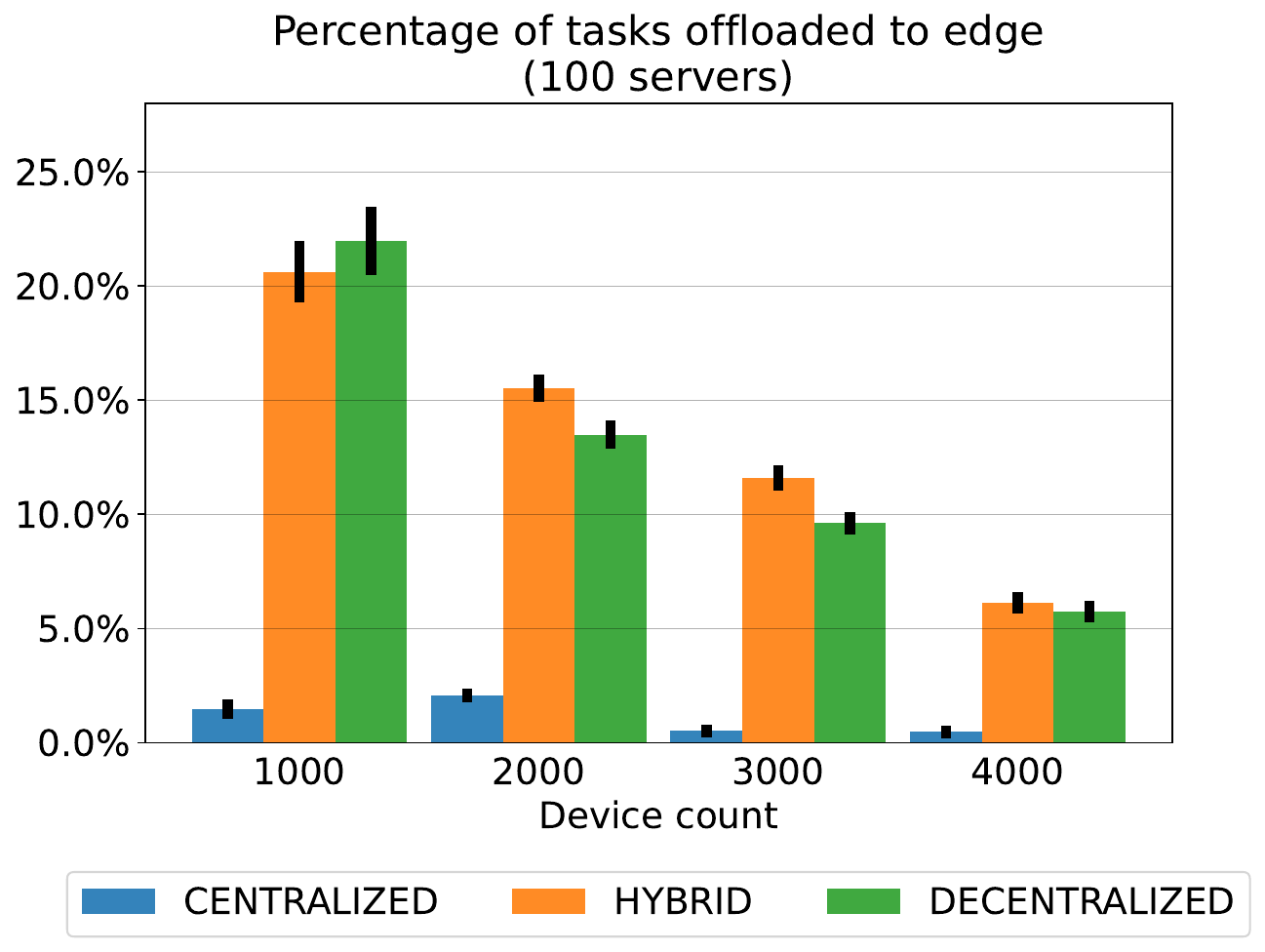}
  \caption{100 servers}
  \label{fig:offloaded100}
\end{subfigure}
\caption{Evaluation plots for the percentage of tasks offloaded to edge.}
\label{fig:offloaded}
\end{figure}

\Cref{fig:offloaded} shows the percentage of tasks offloaded to the edge servers for the 20 high-capacity servers (\Cref{fig:offloaded20}) and the 100 low-capacity servers (\Cref{fig:offloaded100}). Here it can be seen that in all scenarios, the edge devices offload the least amount of tasks in the centralized control topology. This is because in all centralized scenarios, the central orchestrator learns to price at a very high level. In both hybrid and decentralized control topologies, increasing the number of edge devices decreases the percentage of offloaded tasks. This is most likely explained by the pricing strategies of the agents. The examination of the price logs shows that many agents in the environment learn to price a little higher when the edge device count is higher, which also reduces the willingness of the edge devices to offload.

\begin{figure}[!ht]
\centering
\begin{subfigure}{0.49\textwidth}
  \centering
  \includegraphics[width=\linewidth]{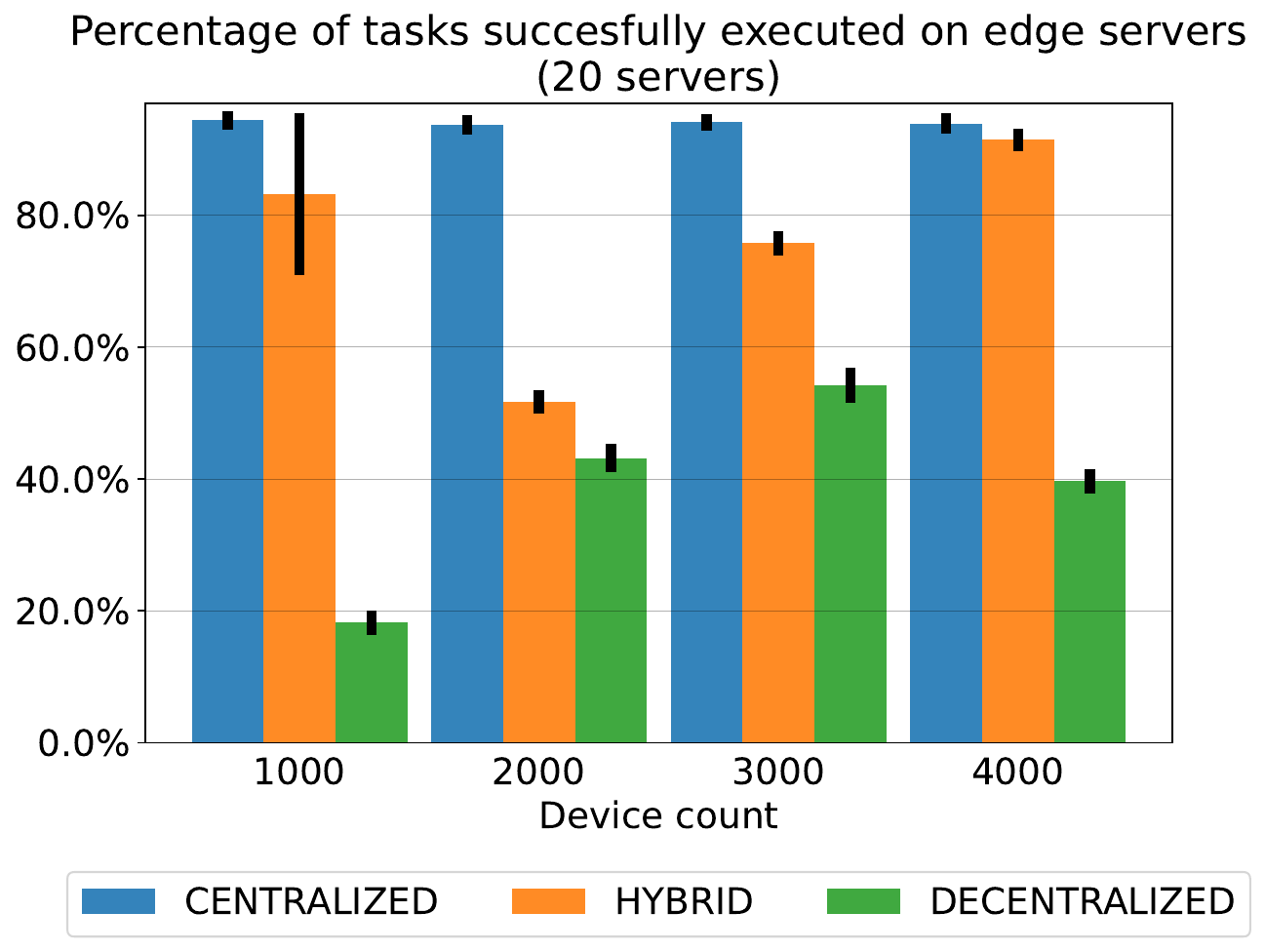}
  \caption{20 servers}
  \label{fig:edgesuccess20}
\end{subfigure}
\begin{subfigure}{0.49\textwidth}
  \centering
  \includegraphics[width=\linewidth]{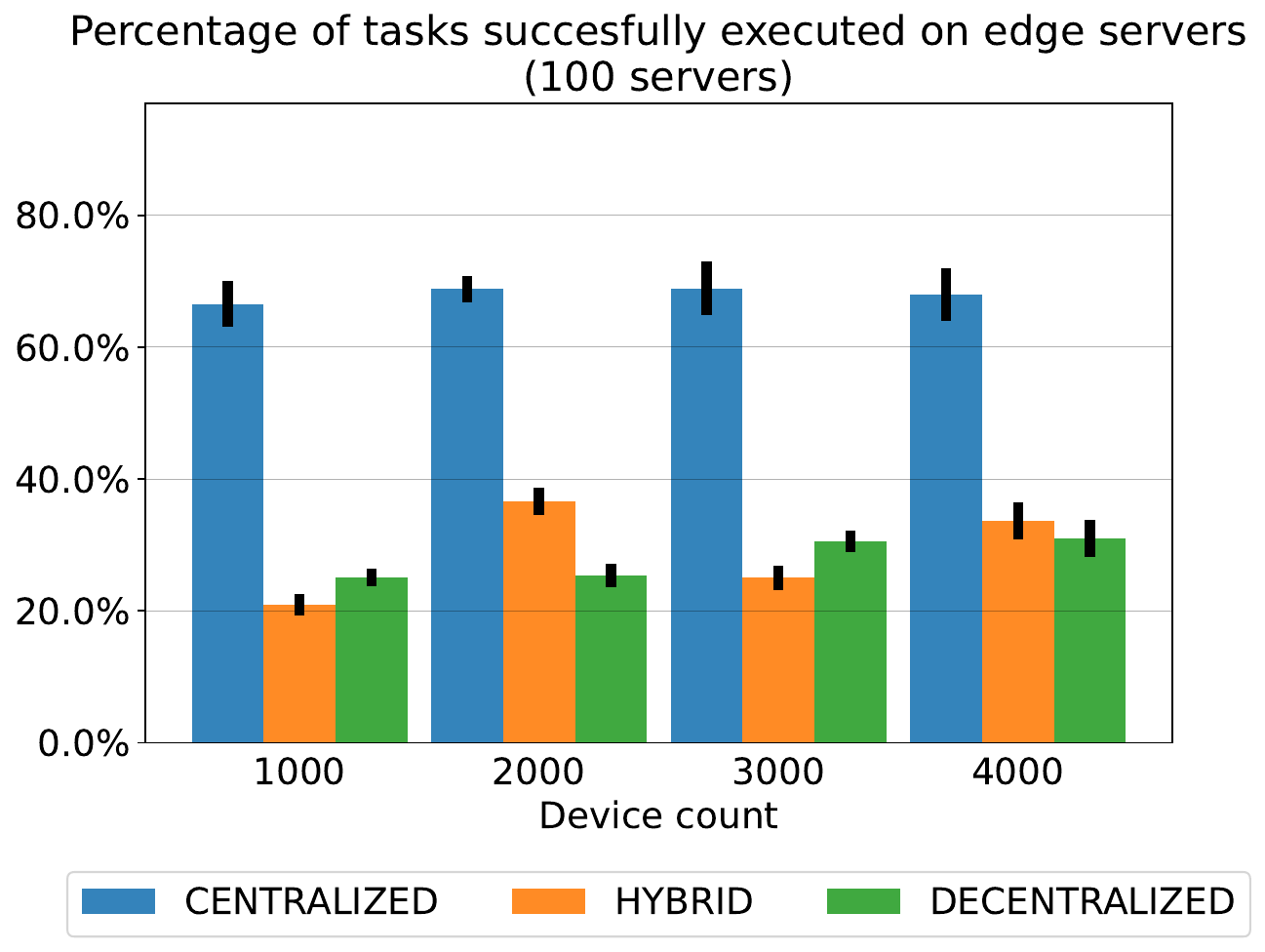}
  \caption{100 servers}
  \label{fig:edgesuccess100}
\end{subfigure}
\caption{Evaluation plots for the percentage of tasks successfully executed on edge servers.}
\label{fig:edgesuccess}
\end{figure}

\Cref{fig:edgesuccess} shows the percentage of the offloaded tasks successfully executed on the edge servers. A task was executed successfully if the latency constraint was satisfied. Here it is good to note that the used constraint is very strict relative to the average task length. In both scenarios with 20 servers (\Cref{fig:edgesuccess20}) and 100 servers (\Cref{fig:edgesuccess100}), the success rate in the centralized control topology is the highest due to the low number of tasks offloaded. Note that the success rate of the centralized topology is lower in the 100-server scenario, which is in line with the lower computational capacity of a single server in the 100-server case. When there are 20 high-capacity servers, hybrid control topology is able to achieve success rates closer to the success rate of the centralized control topology than when there are 100 low-capacity servers. The reduction in the success rate is most likely due to the lower capacity of a single server and the bigger cluster size in the 100-server case.

\begin{figure}[!ht]
\centering
\begin{subfigure}{0.49\textwidth}
  \centering
  \includegraphics[width=\linewidth]{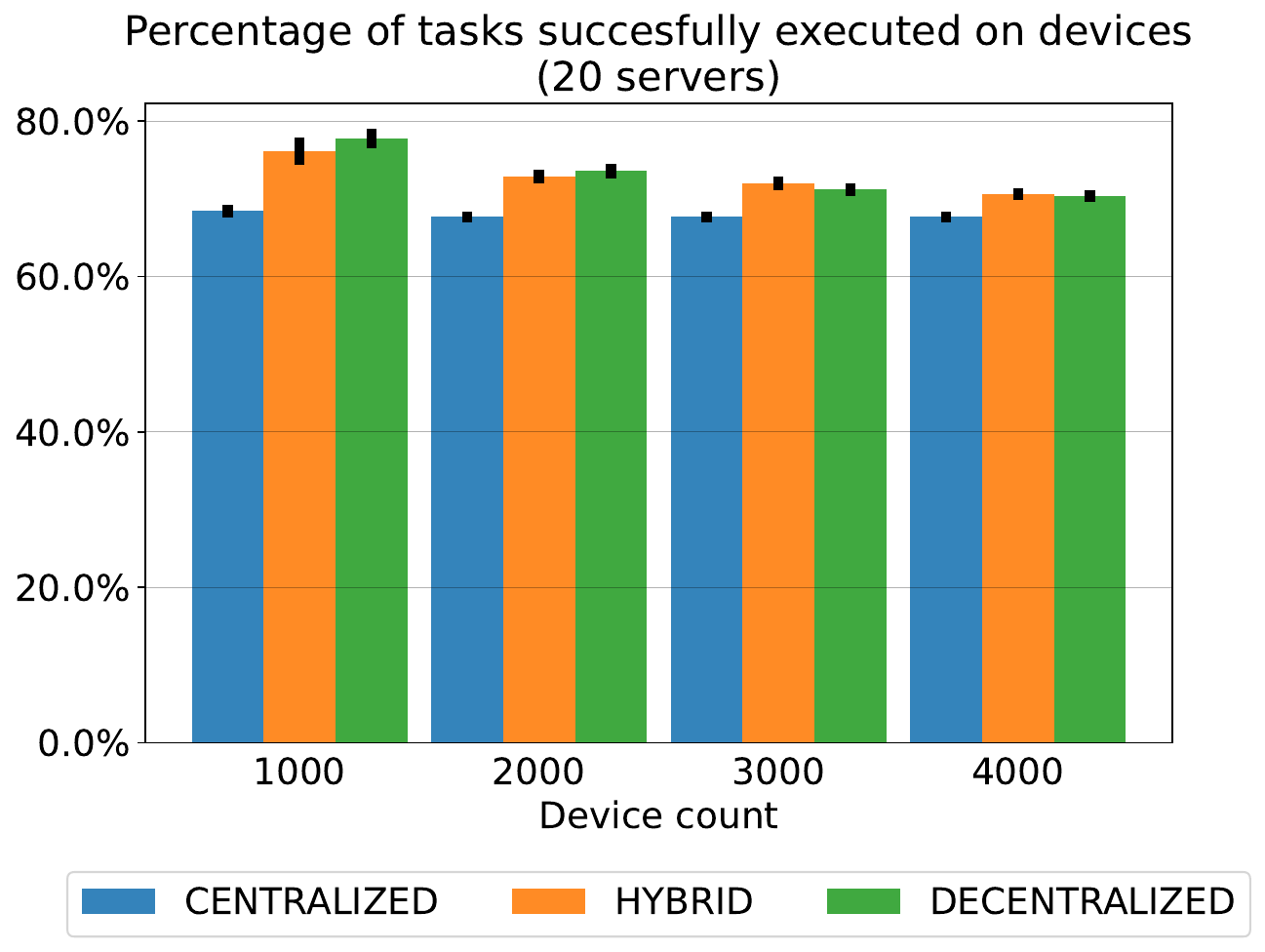}
  \caption{20 servers}
  \label{fig:mistsuccess20}
\end{subfigure}
\begin{subfigure}{0.49\textwidth}
  \centering
  \includegraphics[width=\linewidth]{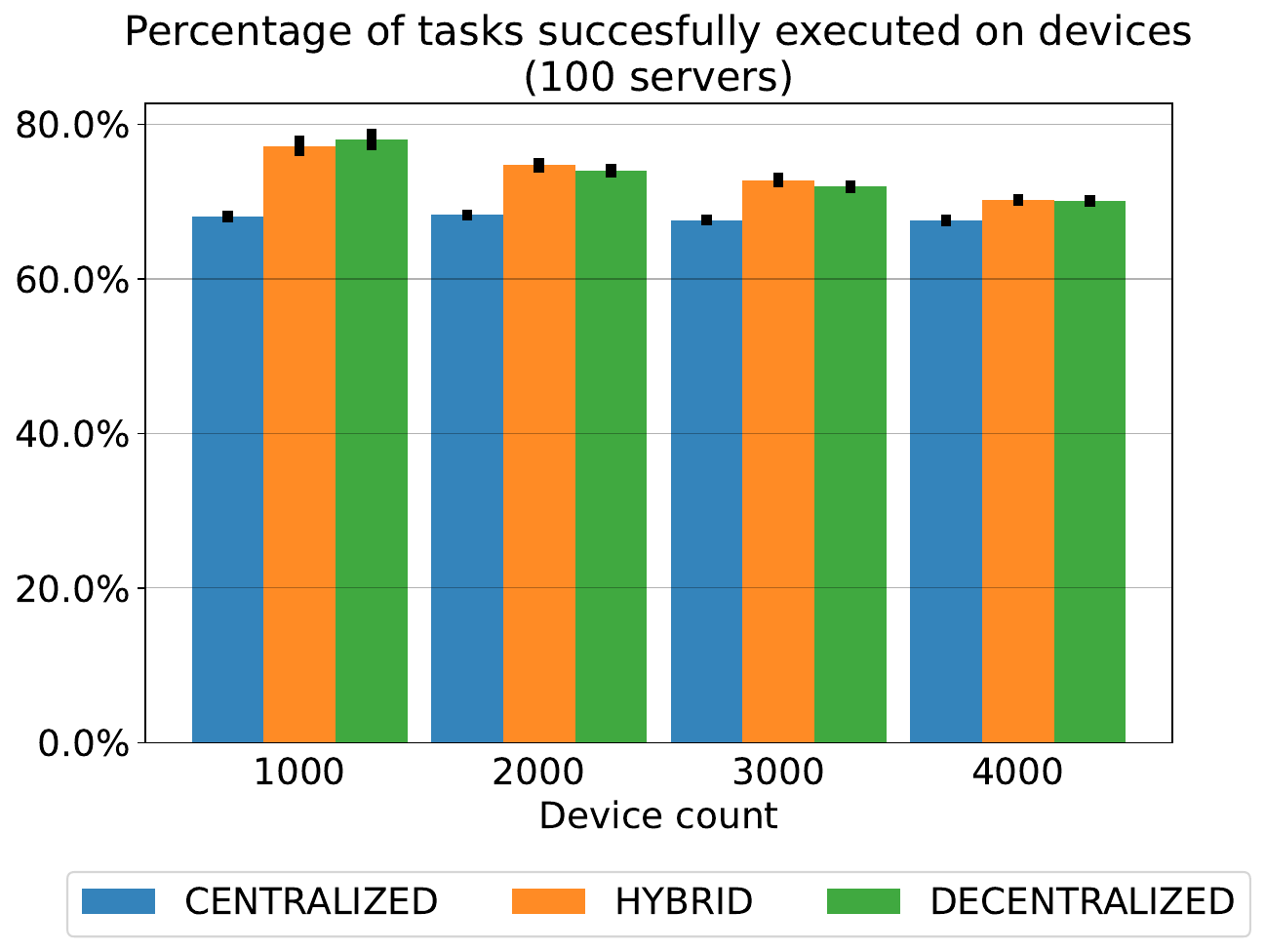}
  \caption{100 servers}
  \label{fig:mistsuccess100}
\end{subfigure}
\caption{Evaluation plots for the percentage of tasks successfully executed on edge devices.}
\label{fig:mistsuccess}
\end{figure}

\Cref{fig:mistsuccess} shows the percentage of the local tasks successfully executed on the edge devices. Here the benefit of the offloading for the edge devices can be seen, as the success rate of the edge devices increases in the hybrid and decentralized control topologies, where more tasks are offloaded when compared to the centralized control topology. This is because devices are more likely to offload lengthy tasks, leaving them with the shorter ones. Further, it is good to note that the increase in the success rate for every scenario is in line with the percentage of offloaded tasks (see \Cref{fig:offloaded}).

\begin{figure}[!ht]
\centering
\begin{subfigure}{0.49\textwidth}
  \centering
  \includegraphics[width=\linewidth]{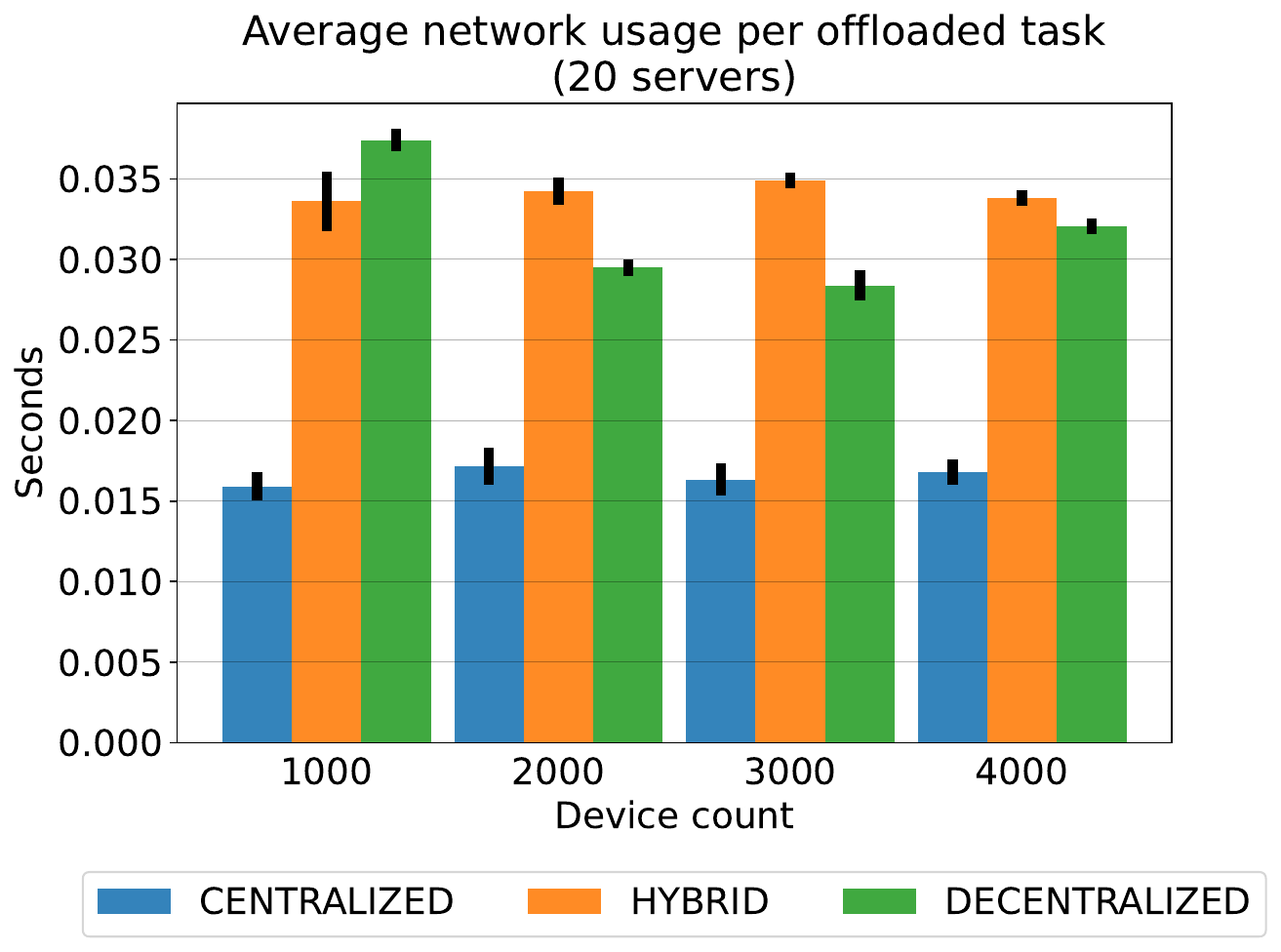}
  \caption{20 servers}
  \label{fig:avgnet20}
\end{subfigure}
\begin{subfigure}{0.49\textwidth}
  \centering
  \includegraphics[width=\linewidth]{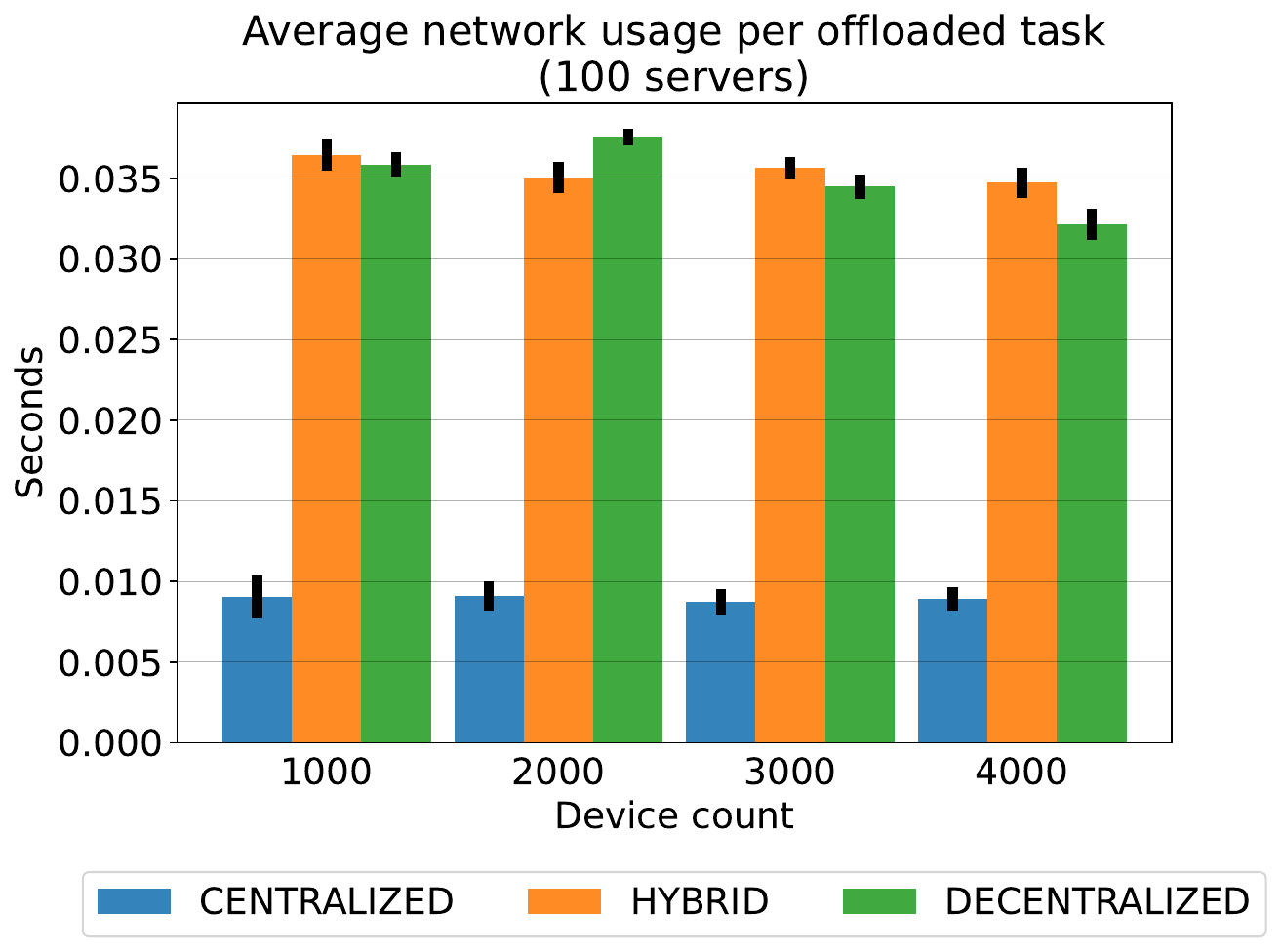}
  \caption{100 servers}
  \label{fig:avgnet100}
\end{subfigure}
\caption{Evaluation plots for the average network usage per offloaded task.}
\label{fig:avgnet}
\end{figure}

\Cref{fig:avgnet} shows the average network usage per offloaded task. The effect of the network congestion on the average can be seen for both edge server counts. As the edge devices do not offload many tasks in the centralized control topology, the average network time per offloaded task is the lowest. Further, the average for the centralized control topology is lower in the 100-server case (\Cref{fig:avgnet100}) than in the 20-server case (\Cref{fig:avgnet20}). This is most likely because the deployment of edge servers is denser in the 100-server case, meaning that the central orchestrator is able to allocate an edge server that is geographically closer to the offloading device than in the 20-server case. In the hybrid and decentralized control topologies, having a denser deployment of edge servers does not cause a similar reduction in the average network time as in the centralized control topology. Particularly in the decentralized control topology, having more servers can increase the average network time. This may reflect the fact that in the 100-server case, a device has more options for offloading in its vicinity, each option with its own price. In other words, the device may choose a server that is not geographically closest to it due to a lower price in a server further away. This increases the distance a task must travel in the network and the congestion in the MAN links.

\begin{figure}[!ht]
\centering
\begin{subfigure}{0.49\textwidth}
  \centering
  \includegraphics[width=\linewidth]{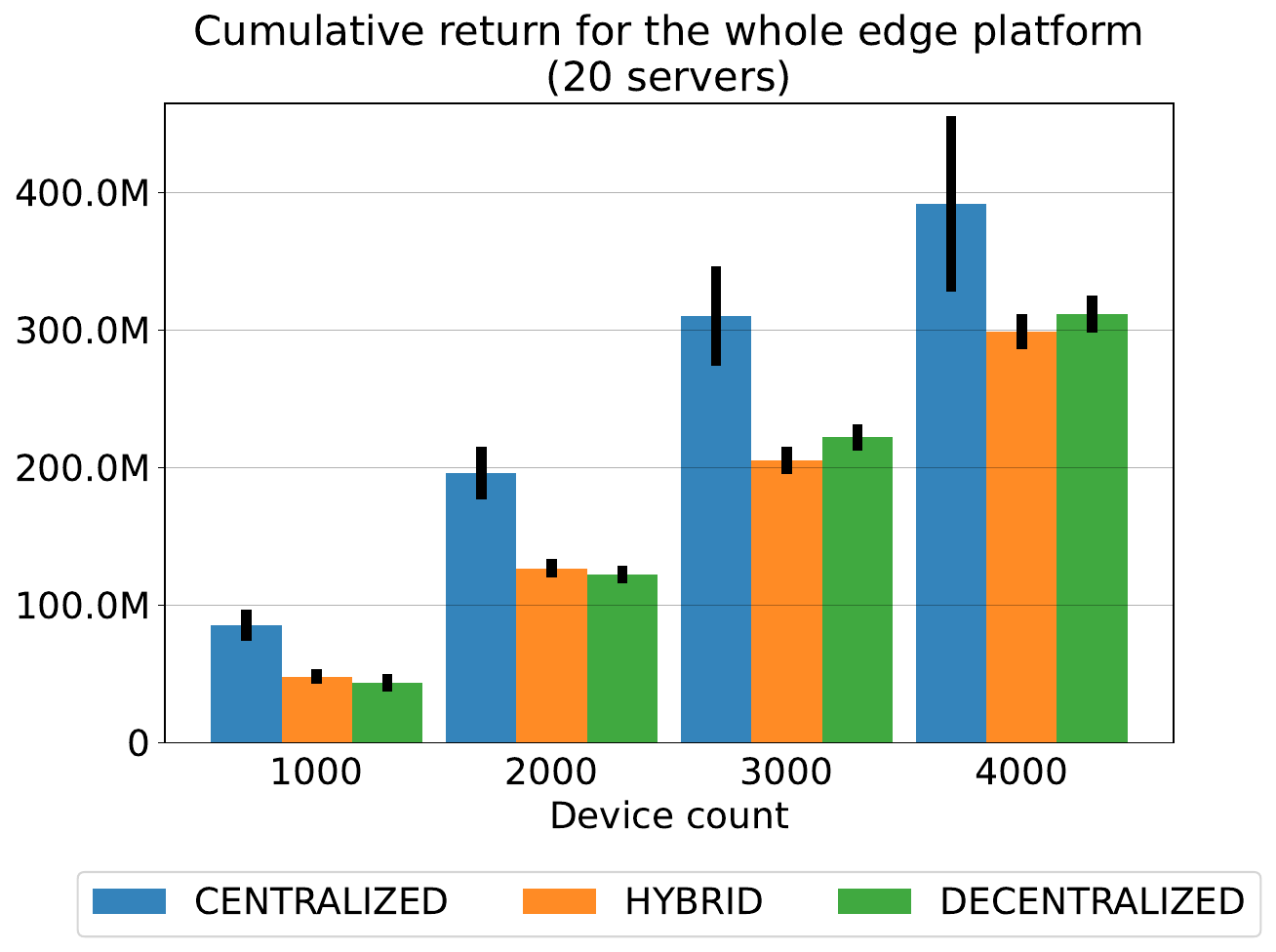}
  \caption{20 servers}
  \label{fig:return20}
\end{subfigure}
\begin{subfigure}{0.49\textwidth}
  \centering
  \includegraphics[width=\linewidth]{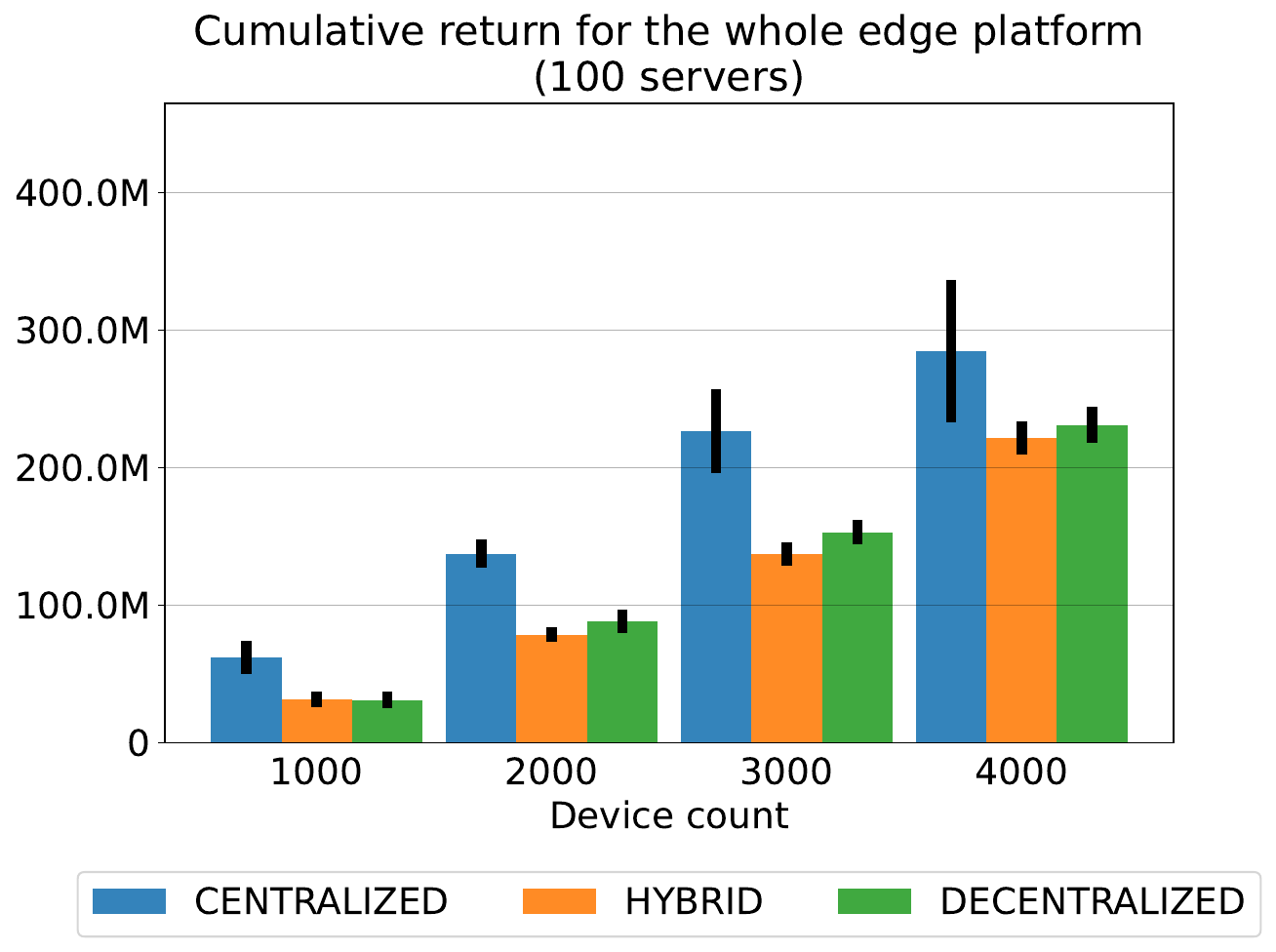}
  \caption{100 servers}
  \label{fig:return100}
\end{subfigure}
\caption{Evaluation plots for the total cumulative return.}
\label{fig:return}
\end{figure}

\Cref{fig:return} shows the total cumulative return for the ESP. The centralized control topology always achieves the highest profit due to the high price set by the central orchestrator. However, the more there are devices in the environment, the more uncertain the profit of the ESP becomes, as reflected in the increased confidence interval for the average. It is also interesting to note that the more there are devices, the closer the profit from the hybrid and decentralized control topologies gets to the centralized one in terms of confidence. In the 100-server case (\Cref{fig:return100}) with 4000 devices, the confidence intervals of all three control topologies overlap, indicating that the hybrid and decentralized control topologies are able to generate more stable profit close to the profit in the centralized control topology with a better resource utilization on the edge platform. Another interesting fact is that in the 20-server case (\Cref{fig:return20}), the profit in every scenario is higher than in the 100-server case. This is most likely due to the lower capacity of a single server in the 100-server case. In other words, edge devices are not willing to pay as much as in the 20-server case, because the benefit of the lower task execution time is not as great as when a server has a higher capacity.

\subsection{Efficiency of EISim}\label{sec:eisim_efficiency}

The time complexity of EISim mainly depends on how many events are generated during a simulation run and how complex the event handling procedures are. To provide insight into the time complexity of EISim, \Cref{tab:sim_times} reports the average time of a single simulation run for each control topology and edge server count combination. The average times are reported separately for the training and evaluation modes. The average and its 95\% confidence interval are calculated based on 100 simulation runs for the training mode and 5 simulation runs for the evaluation mode. It is important to note that one simulation run executed all four scenarios with different edge device counts (1000, 2000, 3000, 4000) in parallel.

\begin{table}[!ht]
\begin{center}
  \caption{Average simulation times and 95\% confidence intervals}
  \label{tab:sim_times}
  \scriptsize
  \begin{tabular}{
    >{\raggedright}p{0.1\linewidth}
    >{\raggedright}p{0.1\linewidth}
    >{\raggedright}p{0.1\linewidth}
    >{\raggedright}p{0.1\linewidth}
    >{\raggedright\arraybackslash}p{0.1\linewidth}
  }
    & \multicolumn{2}{c}{\textbf{Training}} & \multicolumn{2}{c}{\textbf{Evaluation}} \\
    \cmidrule(r){2-3}\cmidrule(r){4-5}
            & \textbf{20 servers} & \textbf{100 servers} & \textbf{20 servers} & \textbf{100 servers} \\
    \midrule 
    \textbf{Centralized} & 7 min 11 s \newline $\pm$ 4 s & 7 min 10 s \newline $\pm$ 5 s & 4 min 38 s \newline $\pm$ 4 s & 4 min 36 s \newline $\pm$ 3 s \\
    \midrule 
    \textbf{Hybrid} & 15 min 4 s \newline $\pm$ 13 s & 21 min 46 s \newline $\pm$ 40 s & 4 min 57 s \newline $\pm$ 17 s & 5 min 10 s \newline $\pm$ 12 s \\
    \midrule 
    \textbf{Decentralized} & 11 min 56 s \newline $\pm$ 4 s & 65 min 8 s \newline $\pm$ 16 s & 4 min 20 s \newline $\pm$ 6 s & 6 min 21 s \newline $\pm$ 4 s \\
    \bottomrule
  \end{tabular}
\end{center}
\end{table}

Two different machines were used to run the simulations. All the simulations for the hybrid control topology and the decentralized control topology with 100 servers were run on a Nokia AirFrame Rackmount server with two Intel Xeon E5-2680 v4 CPUs and 128 GB of RAM. All the simulations for the centralized control topology and the decentralized control topology with 20 servers were run on a desktop computer with Intel Core i7-7800X CPU and 128 GB of RAM. The native CPU backend was used in Deeplearning4j Java library to execute the DNN related computations. It is also important to note is that the machines were running simulations for different scenarios at the same time, which affects the simulation times.

In \Cref{tab:sim_times}, it can be seen that the training of the agents has a significant impact on the average run time. The more there are agents in the environment, the longer it takes to run one training round. On the other hand, having more agents does not significantly prolong the length of an evaluation round.

\subsection{Discussion}

EISim is, as far as is known, the first openly available simulator that specifically supports the simulation of different orchestration control topologies and intelligent, DRL-based orchestration solutions in both single-agent and multi-agent settings. Both of these aspects relate to highly important research questions in the edge orchestration, namely the optimal level of autonomy in orchestration solutions and the adaptation of AI methods in the edge environment. The default implementations for different control topologies provide a good and sensible starting point for research, and they can easily be modified due the high extensibility of EISim.

Resource pricing models for edge environment are an important, largely open research question, but dynamic pricing is generally seen as a key factor in maximizing the profits, optimizing the resource utilization, and creating competition between providers (\cite{Huang2022, Kumar2022}). EISim contributes to this research by allowing the user to simulate dynamic pricing strategies. The default implementation once again provides an excellent starting point that also sets up the foundation for simulating DRL-based solutions. The user can easily modify the existing implementation or plug in a completely new, custom pricing implementation.

The simulation study shows that EISim can provide excellent feedback for choosing hyperparameters (see \Cref{sec:hypertuning}), for investigating the learning behavior of agents and developing training methods (see \Cref{sec:training}), and for comparing the performance of different control topologies (see \Cref{sec:finaleval}). Further, the simulation study shows that EISim can be used to simulate truly large-scale scenarios, as at most the environment had 100 learning agents and 4000 mobile devices.

The main limitations of EISim concern the scope of the simulator and the assumptions made in the default implementations. The current scope of EISim is limited to simulating task offloading with independent tasks and resource pricing. In addition, there is only one pre-implemented DRL algorithm, and the agent model focuses on pricing agents.

The default implementations readily support scenarios where the task processing is done at the edge servers or the edge devices, and the tasks are vertically offloaded from the device level. The servers are also assumed to belong to the same ESP and have the same capacity. Further, the current implementation for the hybrid control topology supports only flat hierarchies with static clusters, which corresponds to the more traditional approach to orchestration control in edge orchestration literature (\cite{Costa2022}).

%% file: Chapters/6_conclusion.tex
A simulation platform called Edge Intelligence Simulator (EISim) was introduced in this article. EISim is developed towards supporting the easier testing and evaluation of intelligent, DRL-based orchestration methods against different orchestration control topologies. In its current form, EISim supports simulating scenarios related to task offloading and resource pricing. 

After providing a detailed description of the architecture, default implementations and use of EISim along with its additional tools for environment setup, agent training, and result plotting, EISim was validated through a large-scale simulation case study. The simulation study verified the end-to-end performance of EISim and showed its capability to produce sensible and consistent results. It was exemplified how the plots generated by EISim can aid in choosing the best hyperparameters, examining the training progress of agents, and comparing the performance of different control topologies.

EISim makes it possible to evaluate the long-term performance of different solutions as a part of a large-scale, more realistic and more dynamic system, and the default implementations provide a solid foundation for further research. EISim can already in its current form provide answers to multiple important, open research questions, such as what type of pricing strategies would be the most efficient for different types of control topologies or how different types of control topologies perform in varying edge deployments and use cases.

\subsection{Future Work}

EISim has a multitude of potential future development directions that it should follow in order to fully live up to its name. These are summarized in the following sections.

\subsubsection{Scope of the Simulator}
For improving the scope of EISim, adding support for other orchestration functions besides offloading and pricing is important. There also needs to be simulation and result analysis support for other types of agents besides pricing agents, such as for RL agents that decide about offloading. To develop the hybrid control topology more towards the vision in \cite{Kokkonen2022}, there must be a better, ready support for multi-level, loosely coupled hierarchies and dynamic clusters. Further, there must be support for modelling multiple service providers in the environment.

Expanding the scope and set of the default implementations is important for having a better, ready support for simulating varying use cases with EISim. Finally, expanding the set of pre-implemented DRL algorithms is also important for facilitating research efforts with EISim.

\subsubsection{Modelling}
For improving the simulation model, it needs to be investigated whether it is beneficial to have a more detailed simulation model, e.g., for networking. Further, to support other orchestration functions such as migration, there needs to be a more detailed model for virtualized resources, such as VMs and containers. Investigating the modelling of security aspects is also important for increasing the realism of the simulation model. Moreover, improving the simulation model into a more dynamic one, where, for example, the time of the day could by simulated by modelling changes in the task arrival rates and the number of edge devices, is another interesting development direction.

Deep neural networks require a significant amount of computing and memory resources. Hence, when it comes to simulating DRL-based solutions, the simulation model should also be able to model how the use of deep neural networks affects the performance of the solution through the resource availability of the system. Investigating efficient ways to model this effect is an important development direction.

\subsubsection{Efficiency and Validation}
The training of agents has a significant impact on the average simulation time, as seen in \Cref{sec:eisim_efficiency} and \Cref{tab:sim_times}. Currently, the updating of the models is tied to an update event, which is handled for each agent sequentially. Parallelizing the model updates is an important development direction for improving the efficiency of EISim.

Further validation of EISim is a necessity for increasing the confidence on the simulator. Hence, future work should also look into building real testbeds for collecting comparable data that could be used to validate EISim. Further validation should also be done by testing EISim over a wider variety of use cases.

\subsubsection{Additional tools}
For improving the scope of the additional tools, future work could look into implementing more methods for MAN creation and clustering. For the result plotting tools, new informative plots could be designed. For example, it could be investigated whether there is an efficient way to visualize the distributions of the prices and state variables for the pricing agents. New types of plots for convergence and training stability analysis should also be designed.

Finally, to improve the ease of use for the additional tools, a graphical user interface could be designed and implemented for the environment setup and result plotting tools.

%% file: main.bbl
\begin{thebibliography}{33}
\expandafter\ifx\csname natexlab\endcsname\relax\def\natexlab#1{#1}\fi
\providecommand{\url}[1]{\texttt{#1}}
\providecommand{\href}[2]{#2}
\providecommand{\path}[1]{#1}
\providecommand{\DOIprefix}{doi:}
\providecommand{\ArXivprefix}{arXiv:}
\providecommand{\URLprefix}{URL: }
\providecommand{\Pubmedprefix}{pmid:}
\providecommand{\doi}[1]{\href{http://dx.doi.org/#1}{\path{#1}}}
\providecommand{\Pubmed}[1]{\href{pmid:#1}{\path{#1}}}
\providecommand{\bibinfo}[2]{#2}
\ifx\xfnm\relax \def\xfnm[#1]{\unskip,\space#1}\fi
\bibitem[{Aral and Maio(2020)}]{Aral2020simulation}
\bibinfo{author}{Aral, A.}, \bibinfo{author}{Maio, V.}, \bibinfo{year}{2020}.
\newblock \bibinfo{title}{Simulators and emulators for edge computing}, in: \bibinfo{editor}{Taheri, J.}, \bibinfo{editor}{Deng, S.} (Eds.), \bibinfo{booktitle}{Edge Computing: Models, technologies and applications}, pp. \bibinfo{pages}{291--311}.
\newblock \DOIprefix\doi{10.1049/PBPC033E\_ch14}.
\bibitem[{Calheiros et~al.(2011)Calheiros, Ranjan, Beloglazov, De~Rose and Buyya}]{cloudsim}
\bibinfo{author}{Calheiros, R.N.}, \bibinfo{author}{Ranjan, R.}, \bibinfo{author}{Beloglazov, A.}, \bibinfo{author}{De~Rose, C.A.F.}, \bibinfo{author}{Buyya, R.}, \bibinfo{year}{2011}.
\newblock \bibinfo{title}{{{CloudSim}}: A toolkit for modeling and simulation of cloud computing environments and evaluation of resource provisioning algorithms}.
\newblock \bibinfo{journal}{Software: Practice and Experience} \bibinfo{volume}{41(1)}, \bibinfo{pages}{23--50}.
\newblock \DOIprefix\doi{10.1002/spe.995}.
\bibitem[{Costa et~al.(2022)Costa, Bachiega, de~Carvalho and Araujo}]{Costa2022}
\bibinfo{author}{Costa, B.}, \bibinfo{author}{Bachiega, J.}, \bibinfo{author}{de~Carvalho, L.R.}, \bibinfo{author}{Araujo, A.P.F.}, \bibinfo{year}{2022}.
\newblock \bibinfo{title}{Orchestration in fog computing: A comprehensive survey}.
\newblock \bibinfo{journal}{ACM Computing Surveys} \bibinfo{volume}{55(2)}, \bibinfo{pages}{1--34}.
\newblock \DOIprefix\doi{10.1145/3486221}.
\bibitem[{Deng et~al.(2020)Deng, Zhao, Fang, Yin, Dustdar and Zomaya}]{Deng2019}
\bibinfo{author}{Deng, S.}, \bibinfo{author}{Zhao, H.}, \bibinfo{author}{Fang, W.}, \bibinfo{author}{Yin, J.}, \bibinfo{author}{Dustdar, S.}, \bibinfo{author}{Zomaya, A.Y.}, \bibinfo{year}{2020}.
\newblock \bibinfo{title}{Edge intelligence: The confluence of edge computing and artificial intelligence}.
\newblock \bibinfo{journal}{IEEE Internet of Things Journal} \bibinfo{volume}{7(8)}, \bibinfo{pages}{7457--7469}.
\newblock \DOIprefix\doi{10.1109/JIOT.2020.2984887}.
\bibitem[{Fernández-Cerero et~al.(2020)Fernández-Cerero, Fernández-Montes, Ortega, Jakóbik and Widlak}]{sphere}
\bibinfo{author}{Fernández-Cerero, D.}, \bibinfo{author}{Fernández-Montes, A.}, \bibinfo{author}{Ortega, F.J.}, \bibinfo{author}{Jakóbik, A.}, \bibinfo{author}{Widlak, A.}, \bibinfo{year}{2020}.
\newblock \bibinfo{title}{Sphere: Simulator of edge infrastructures for the optimization of performance and resources energy consumption}.
\newblock \bibinfo{journal}{Simulation Modelling Practice and Theory} \bibinfo{volume}{101}.
\newblock \DOIprefix\doi{10.1016/j.simpat.2019.101966}.
\bibitem[{Gilchrist(2016)}]{gilchrist2016}
\bibinfo{author}{Gilchrist, A.}, \bibinfo{year}{2016}.
\newblock \bibinfo{title}{Industry 4.0}.
\newblock \bibinfo{publisher}{Apress Berkeley, CA}.
\newblock \DOIprefix\doi{10.1007/978-1-4842-2047-4}.
\bibitem[{Gill and Singh(2021)}]{Gill2021simulation}
\bibinfo{author}{Gill, M.}, \bibinfo{author}{Singh, D.}, \bibinfo{year}{2021}.
\newblock \bibinfo{title}{A comprehensive study of simulation frameworks and research directions in fog computing}.
\newblock \bibinfo{journal}{Computer Science Review} \bibinfo{volume}{40}.
\newblock \DOIprefix\doi{10.1016/j.cosrev.2021.100391}.
\bibitem[{Gupta et~al.(2017)Gupta, Dastjerdi, Ghosh and Buyya}]{ifogsim}
\bibinfo{author}{Gupta, H.}, \bibinfo{author}{Dastjerdi, A.V.}, \bibinfo{author}{Ghosh, S.K.}, \bibinfo{author}{Buyya, R.}, \bibinfo{year}{2017}.
\newblock \bibinfo{title}{{{iFogSim}}: A toolkit for modeling and simulation of resource management techniques in the internet of things, edge and fog computing environments}.
\newblock \bibinfo{journal}{Software: Practice and Experience} \bibinfo{volume}{47(9)}, \bibinfo{pages}{1275--1296}.
\newblock \DOIprefix\doi{10.1002/spe.2509}.
\bibitem[{Huang et~al.(2022)Huang, Zhang and Li}]{Huang2022}
\bibinfo{author}{Huang, X.}, \bibinfo{author}{Zhang, B.}, \bibinfo{author}{Li, C.}, \bibinfo{year}{2022}.
\newblock \bibinfo{title}{Incentive mechanisms for mobile edge computing: Present and future directions}.
\newblock \bibinfo{journal}{IEEE Network} \bibinfo{volume}{36(6)}, \bibinfo{pages}{199--205}.
\newblock \DOIprefix\doi{10.1109/MNET.107.2100652}.
\bibitem[{Kokkonen et~al.(2022)Kokkonen, Lovén, Motlagh, Partala, González-Gil, Sola, Angulo, Liyanage, Leppänen, Nguyen, Pujol, Kostakos, Bennis, Tarkoma, Dustdar, Pirttikangas and Riekki}]{Kokkonen2022}
\bibinfo{author}{Kokkonen, H.}, \bibinfo{author}{Lovén, L.}, \bibinfo{author}{Motlagh, N.H.}, \bibinfo{author}{Partala, J.}, \bibinfo{author}{González-Gil, A.}, \bibinfo{author}{Sola, E.}, \bibinfo{author}{Angulo, I.}, \bibinfo{author}{Liyanage, M.}, \bibinfo{author}{Leppänen, T.}, \bibinfo{author}{Nguyen, T.}, \bibinfo{author}{Pujol, V.C.}, \bibinfo{author}{Kostakos, P.}, \bibinfo{author}{Bennis, M.}, \bibinfo{author}{Tarkoma, S.}, \bibinfo{author}{Dustdar, S.}, \bibinfo{author}{Pirttikangas, S.}, \bibinfo{author}{Riekki, J.}, \bibinfo{year}{2022}.
\newblock \bibinfo{title}{Autonomy and intelligence in the computing continuum: Challenges, enablers, and future directions for orchestration}.
\newblock \bibinfo{howpublished}{arXiv:2205.01423}.
\newblock \DOIprefix\doi{10.48550/ARXIV.2205.01423}.
\bibitem[{Kumar et~al.(2022)Kumar, Baranwal and Vidyarthi}]{Kumar2022}
\bibinfo{author}{Kumar, D.}, \bibinfo{author}{Baranwal, G.}, \bibinfo{author}{Vidyarthi, D.P.}, \bibinfo{year}{2022}.
\newblock \bibinfo{title}{A survey on auction based approaches for resource allocation and pricing in emerging edge technologies}.
\newblock \bibinfo{journal}{Journal of Grid Computing} \bibinfo{volume}{20(1)}.
\newblock \DOIprefix\doi{10.1007/s10723-021-09593-9}.
\bibitem[{Lera et~al.(2019)Lera, Guerrero and Juiz}]{yafs}
\bibinfo{author}{Lera, I.}, \bibinfo{author}{Guerrero, C.}, \bibinfo{author}{Juiz, C.}, \bibinfo{year}{2019}.
\newblock \bibinfo{title}{{{YAFS}}: A simulator for iot scenarios in fog computing}.
\newblock \bibinfo{journal}{IEEE Access} \bibinfo{volume}{7}, \bibinfo{pages}{91745--91758}.
\newblock \DOIprefix\doi{10.1109/ACCESS.2019.2927895}.
\bibitem[{Lillicrap et~al.(2019)Lillicrap, Hunt, Pritzel, Heess, Erez, Tassa, Silver and Wierstra}]{ddpg}
\bibinfo{author}{Lillicrap, T.P.}, \bibinfo{author}{Hunt, J.J.}, \bibinfo{author}{Pritzel, A.}, \bibinfo{author}{Heess, N.}, \bibinfo{author}{Erez, T.}, \bibinfo{author}{Tassa, Y.}, \bibinfo{author}{Silver, D.}, \bibinfo{author}{Wierstra, D.}, \bibinfo{year}{2019}.
\newblock \bibinfo{title}{Continuous control with deep reinforcement learning}.
\newblock \bibinfo{howpublished}{arXiv:1509.02971}.
\newblock \DOIprefix\doi{10.48550/arXiv.1509.02971}.
\bibitem[{Lov{\'{e}}n et~al.(2019)Lov{\'{e}}n, Lepp{\"{a}}nen, Peltonen, Partala, Harjula, Porambage, Ylianttila and Riekki}]{Loven2019}
\bibinfo{author}{Lov{\'{e}}n, L.}, \bibinfo{author}{Lepp{\"{a}}nen, T.}, \bibinfo{author}{Peltonen, E.}, \bibinfo{author}{Partala, J.}, \bibinfo{author}{Harjula, E.}, \bibinfo{author}{Porambage, P.}, \bibinfo{author}{Ylianttila, M.}, \bibinfo{author}{Riekki, J.}, \bibinfo{year}{2019}.
\newblock \bibinfo{title}{{EdgeAI: A vision for distributed, edge-native artificial intelligence in future 6G networks}}, in: \bibinfo{booktitle}{The 1st 6G Wireless Summit}, \bibinfo{address}{Levi, Finland}. pp. \bibinfo{pages}{1--2}.
\bibitem[{Lov\'{e}n et~al.(2022)Lov\'{e}n, Peltonen, Ruha, Harjula and Pirttikangas}]{Loven2022}
\bibinfo{author}{Lov\'{e}n, L.}, \bibinfo{author}{Peltonen, E.}, \bibinfo{author}{Ruha, L.}, \bibinfo{author}{Harjula, E.}, \bibinfo{author}{Pirttikangas, S.}, \bibinfo{year}{2022}.
\newblock \bibinfo{title}{A dark and stormy night: Reallocation storms in edge computing}.
\newblock \bibinfo{journal}{EURASIP Journal on Wireless Communications and Networking} \bibinfo{volume}{2022(1)}.
\newblock \DOIprefix\doi{10.1186/s13638-022-02170-y}.
\bibitem[{Lähderanta et~al.(2021)Lähderanta, Leppänen, Ruha, Lovén, Harjula, Ylianttila, Riekki and Sillanpää}]{Lahderanta2021}
\bibinfo{author}{Lähderanta, T.}, \bibinfo{author}{Leppänen, T.}, \bibinfo{author}{Ruha, L.}, \bibinfo{author}{Lovén, L.}, \bibinfo{author}{Harjula, E.}, \bibinfo{author}{Ylianttila, M.}, \bibinfo{author}{Riekki, J.}, \bibinfo{author}{Sillanpää, M.J.}, \bibinfo{year}{2021}.
\newblock \bibinfo{title}{Edge computing server placement with capacitated location allocation}.
\newblock \bibinfo{journal}{Journal of Parallel and Distributed Computing} \bibinfo{volume}{153}, \bibinfo{pages}{130--149}.
\newblock \DOIprefix\doi{10.1016/j.jpdc.2021.03.007}.
\bibitem[{Mahmud et~al.(2022)Mahmud, Pallewatta, Goudarzi and Buyya}]{ifogsim2}
\bibinfo{author}{Mahmud, R.}, \bibinfo{author}{Pallewatta, S.}, \bibinfo{author}{Goudarzi, M.}, \bibinfo{author}{Buyya, R.}, \bibinfo{year}{2022}.
\newblock \bibinfo{title}{{{iFogSim2}}: An extended {{iFogSim}} simulator for mobility, clustering, and microservice management in edge and fog computing environments}.
\newblock \bibinfo{journal}{Journal of Systems and Software} \bibinfo{volume}{190}.
\newblock \DOIprefix\doi{10.1016/j.jss.2022.111351}.
\bibitem[{Mechalikh et~al.(2021)Mechalikh, Taktak and Moussa}]{pureEdgeSim}
\bibinfo{author}{Mechalikh, C.}, \bibinfo{author}{Taktak, H.}, \bibinfo{author}{Moussa, F.}, \bibinfo{year}{2021}.
\newblock \bibinfo{title}{{PureEdgeSim: A simulation framework for performance evaluation of cloud, edge and mist computing environments}}.
\newblock \bibinfo{journal}{Computer Science and Information Systems} \bibinfo{volume}{18(1)}, \bibinfo{pages}{43--66}.
\newblock \DOIprefix\doi{10.2298/CSIS200301042M}.
\bibitem[{Mämmelä and Riekki(2021)}]{Mammela2021}
\bibinfo{author}{Mämmelä, A.}, \bibinfo{author}{Riekki, J.}, \bibinfo{year}{2021}.
\newblock \bibinfo{title}{Subsidiarity and weak coupling in wireless networks}, in: \bibinfo{booktitle}{2021 Joint European Conference on Networks and Communications 6G Summit (EuCNC/6G Summit)}, \bibinfo{address}{Porto, Portugal}. pp. \bibinfo{pages}{598--603}.
\newblock \DOIprefix\doi{10.1109/EuCNC/6GSummit51104.2021.9482591}.
\bibitem[{Park et~al.(2021)Park, Samarakoon, Elgabli, Kim, Bennis, Kim and Debbah}]{Park2021}
\bibinfo{author}{Park, J.}, \bibinfo{author}{Samarakoon, S.}, \bibinfo{author}{Elgabli, A.}, \bibinfo{author}{Kim, J.}, \bibinfo{author}{Bennis, M.}, \bibinfo{author}{Kim, S.L.}, \bibinfo{author}{Debbah, M.}, \bibinfo{year}{2021}.
\newblock \bibinfo{title}{Communication-efficient and distributed learning over wireless networks: Principles and applications}.
\newblock \bibinfo{journal}{Proceedings of the IEEE} \bibinfo{volume}{109(5)}, \bibinfo{pages}{796--819}.
\newblock \DOIprefix\doi{10.1109/JPROC.2021.3055679}.
\bibitem[{Puliafito et~al.(2020)Puliafito, Gonçalves, Lopes, Martins, Madeira, Mingozzi, Rana and Bittencourt}]{mobfogsim}
\bibinfo{author}{Puliafito, C.}, \bibinfo{author}{Gonçalves, D.M.}, \bibinfo{author}{Lopes, M.M.}, \bibinfo{author}{Martins, L.L.}, \bibinfo{author}{Madeira, E.}, \bibinfo{author}{Mingozzi, E.}, \bibinfo{author}{Rana, O.}, \bibinfo{author}{Bittencourt, L.F.}, \bibinfo{year}{2020}.
\newblock \bibinfo{title}{{{MobFogSim}}: Simulation of mobility and migration for fog computing}.
\newblock \bibinfo{journal}{Simulation Modelling Practice and Theory} \bibinfo{volume}{101}.
\newblock \DOIprefix\doi{10.1016/j.simpat.2019.102062}.
\bibitem[{Qadri et~al.(2020)Qadri, Nauman, Zikria, Vasilakos and Kim}]{qadri2020}
\bibinfo{author}{Qadri, Y.A.}, \bibinfo{author}{Nauman, A.}, \bibinfo{author}{Zikria, Y.B.}, \bibinfo{author}{Vasilakos, A.V.}, \bibinfo{author}{Kim, S.W.}, \bibinfo{year}{2020}.
\newblock \bibinfo{title}{The future of healthcare internet of things: A survey of emerging technologies}.
\newblock \bibinfo{journal}{IEEE Communications Surveys {\&} Tutorials} \bibinfo{volume}{22(2)}, \bibinfo{pages}{1121--1167}.
\newblock \DOIprefix\doi{10.1109/COMST.2020.2973314}.
\bibitem[{Qayyum et~al.(2018)Qayyum, Malik, Khan~Khattak, Khalid and Khan}]{fognetsim}
\bibinfo{author}{Qayyum, T.}, \bibinfo{author}{Malik, A.W.}, \bibinfo{author}{Khan~Khattak, M.A.}, \bibinfo{author}{Khalid, O.}, \bibinfo{author}{Khan, S.U.}, \bibinfo{year}{2018}.
\newblock \bibinfo{title}{{{FogNetSim++}}: A toolkit for modeling and simulation of distributed fog environment}.
\newblock \bibinfo{journal}{IEEE Access} \bibinfo{volume}{6}, \bibinfo{pages}{63570--63583}.
\newblock \DOIprefix\doi{10.1109/ACCESS.2018.2877696}.
\bibitem[{Ren et~al.(2020)Ren, Zhang, He, Zhang and Li}]{Ren2020}
\bibinfo{author}{Ren, J.}, \bibinfo{author}{Zhang, D.}, \bibinfo{author}{He, S.}, \bibinfo{author}{Zhang, Y.}, \bibinfo{author}{Li, T.}, \bibinfo{year}{2020}.
\newblock \bibinfo{title}{A survey on end-edge-cloud orchestrated network computing paradigms: Transparent computing, mobile edge computing, fog computing, and cloudlet}.
\newblock \bibinfo{journal}{ACM Computing Surveys} \bibinfo{volume}{52(6)}, \bibinfo{pages}{1--36}.
\newblock \DOIprefix\doi{10.1145/3362031}.
\bibitem[{{Saraiva de Sousa} et~al.(2019){Saraiva de Sousa}, {Lachos Perez}, Rosa, Santos and {Esteve Rothenberg}}]{de2019network}
\bibinfo{author}{{Saraiva de Sousa}, N.F.}, \bibinfo{author}{{Lachos Perez}, D.A.}, \bibinfo{author}{Rosa, R.V.}, \bibinfo{author}{Santos, M.A.}, \bibinfo{author}{{Esteve Rothenberg}, C.}, \bibinfo{year}{2019}.
\newblock \bibinfo{title}{Network service orchestration: A survey}.
\newblock \bibinfo{journal}{Computer Communications} \bibinfo{volume}{142-143}, \bibinfo{pages}{69--94}.
\newblock \DOIprefix\doi{10.1016/j.comcom.2019.04.008}.
\bibitem[{Shaik et~al.(2022)Shaik, Hall, Johnson, Wang, Sharp and Baskiyar}]{pfogsim}
\bibinfo{author}{Shaik, S.}, \bibinfo{author}{Hall, J.}, \bibinfo{author}{Johnson, C.}, \bibinfo{author}{Wang, Q.}, \bibinfo{author}{Sharp, R.}, \bibinfo{author}{Baskiyar, S.}, \bibinfo{year}{2022}.
\newblock \bibinfo{title}{{{PFogSim}}: A simulator for evaluation of mobile and hierarchical fog computing}.
\newblock \bibinfo{journal}{Sustainable Computing: Informatics and Systems} \bibinfo{volume}{35}.
\newblock \DOIprefix\doi{10.1016/j.suscom.2022.100736}.
\bibitem[{Soltan and Zussman(2016)}]{Soltan2016}
\bibinfo{author}{Soltan, S.}, \bibinfo{author}{Zussman, G.}, \bibinfo{year}{2016}.
\newblock \bibinfo{title}{Generation of synthetic spatially embedded power grid networks}, in: \bibinfo{booktitle}{2016 {{IEEE Power}} and {{Energy Society General Meeting}} ({{PESGM}})}, \bibinfo{address}{Boston, MA, USA}. pp. \bibinfo{pages}{1--5}.
\newblock \DOIprefix\doi{10.1109/PESGM.2016.7741383}.
\bibitem[{Sonmez et~al.(2018)Sonmez, Ozgovde and Ersoy}]{edgecloudsim}
\bibinfo{author}{Sonmez, C.}, \bibinfo{author}{Ozgovde, A.}, \bibinfo{author}{Ersoy, C.}, \bibinfo{year}{2018}.
\newblock \bibinfo{title}{{{EdgeCloudSim}}: An environment for performance evaluation of edge computing systems}.
\newblock \bibinfo{journal}{Transactions on Emerging Telecommunications Technologies} \bibinfo{volume}{29(11)}.
\newblock \DOIprefix\doi{10.1002/ett.3493}.
\bibitem[{Taleb et~al.(2017)Taleb, Samdanis, Mada, Flinck, Dutta and Sabella}]{Taleb2017}
\bibinfo{author}{Taleb, T.}, \bibinfo{author}{Samdanis, K.}, \bibinfo{author}{Mada, B.}, \bibinfo{author}{Flinck, H.}, \bibinfo{author}{Dutta, S.}, \bibinfo{author}{Sabella, D.}, \bibinfo{year}{2017}.
\newblock \bibinfo{title}{On multi-access edge computing: A survey of the emerging {{5G}} network edge cloud architecture and orchestration}.
\newblock \bibinfo{journal}{IEEE Communications Surveys \& Tutorials} \bibinfo{volume}{19(3)}, \bibinfo{pages}{1657--1681}.
\newblock \DOIprefix\doi{10.1109/COMST.2017.2705720}.
\bibitem[{Varga and Hornig(2008)}]{omnet}
\bibinfo{author}{Varga, A.}, \bibinfo{author}{Hornig, R.}, \bibinfo{year}{2008}.
\newblock \bibinfo{title}{An overview of the {{OMNeT++}} simulation environment}, in: \bibinfo{booktitle}{Proceedings of the 1st International Conference on Simulation Tools and Techniques for Communications, Networks and Systems \& Workshops (Simutools '08)}, \bibinfo{publisher}{ICST (Institute for Computer Sciences, Social-Informatics and Telecommunications Engineering)}, \bibinfo{address}{Brussels, BEL}. pp. \bibinfo{pages}{1--10}.
\bibitem[{Wang et~al.(2021)Wang, Li, Li, Qiu and Wang}]{SimEdgeIntel}
\bibinfo{author}{Wang, C.}, \bibinfo{author}{Li, R.}, \bibinfo{author}{Li, W.}, \bibinfo{author}{Qiu, C.}, \bibinfo{author}{Wang, X.}, \bibinfo{year}{2021}.
\newblock \bibinfo{title}{Simedgeintel: A open-source simulation platform for resource management in edge intelligence}.
\newblock \bibinfo{journal}{Journal of Systems Architecture} \bibinfo{volume}{115}.
\newblock \DOIprefix\doi{10.1016/j.sysarc.2021.102016}.
\bibitem[{Xu et~al.(2021)Xu, Li, Li, Su, Tarkoma, Jiang, Crowcroft and Hui}]{Xu2020}
\bibinfo{author}{Xu, D.}, \bibinfo{author}{Li, T.}, \bibinfo{author}{Li, Y.}, \bibinfo{author}{Su, X.}, \bibinfo{author}{Tarkoma, S.}, \bibinfo{author}{Jiang, T.}, \bibinfo{author}{Crowcroft, J.}, \bibinfo{author}{Hui, P.}, \bibinfo{year}{2021}.
\newblock \bibinfo{title}{Edge intelligence: Empowering intelligence to the edge of network}.
\newblock \bibinfo{journal}{Proceedings of the IEEE} \bibinfo{volume}{109(11)}, \bibinfo{pages}{1778--1837}.
\newblock \DOIprefix\doi{10.1109/JPROC.2021.3119950}.
\bibitem[{Zhong et~al.(2022)Zhong, Xu, Rodriguez, Xu and Buyya}]{Zhong2022}
\bibinfo{author}{Zhong, Z.}, \bibinfo{author}{Xu, M.}, \bibinfo{author}{Rodriguez, M.A.}, \bibinfo{author}{Xu, C.}, \bibinfo{author}{Buyya, R.}, \bibinfo{year}{2022}.
\newblock \bibinfo{title}{Machine learning-based orchestration of containers: A taxonomy and future directions}.
\newblock \bibinfo{journal}{ACM Computing Surveys} \bibinfo{volume}{54(10s)}.
\newblock \DOIprefix\doi{10.1145/3510415}.

\end{thebibliography}
